\documentclass[iop]{emulateapj-rtx4}
\usepackage{graphicx,wrapfig}
\usepackage{booktabs}
\usepackage{amssymb}
\usepackage{enumitem}
\usepackage{amsmath}
\usepackage{amssymb}
\usepackage{ctable}
\usepackage{color}
\usepackage{afterpage}
\usepackage{soul}
\usepackage{threeparttable}

\newcommand{\sgra}{Sgr~A*}

\newcommand{\abh}{\alpha_{\mbox{\scriptsize bh}}}
\newcommand{\apsr}{\alpha_{\mbox{\scriptsize psr}}}

\newcommand{\Ssgr}{S_{\mbox{\scriptsize Sgr}}}
\newcommand{\Sbhvo}{S_{\mbox{\scriptsize BH,$\nu_0$}}}
\newcommand{\Spsrvo}{S_{\mbox{\scriptsize psr,$\nu_0$}}}
\newcommand{\Spsronefour}{S_{\mbox{\scriptsize psr,$1.4$}}}
\newcommand{\be}{\begin{eqnarray}}
\newcommand{\ee}{\end{eqnarray}}
\newcommand{\thetasc}{\theta_{\rm sc}}
\newcommand{\thetascz}{\theta_{\rm sc_0}}
\newcommand{\thetab}{\theta_{\rm b}}
\newcommand{\thetabz}{\theta_{\rm b_0}}
\newcommand{\Omegaeff}{\Omega_{\rm eff}}
\newcommand{\arcs}{{}^{\prime\prime}}

\newcommand{\h}{{}^{\mbox{\scriptsize h}}}
\newcommand{\m}{{}^{\mbox{\scriptsize m}}}

\newcommand{\ncp}{N_{\mbox{\scriptsize {CP}}}}
\newcommand{\nmsp}{N_{\mbox{\scriptsize MSP}}}
\newcommand{\nughz}{\nu_{\mbox{\scriptsize GHz}}}

\begin{document}
\shorttitle{Constraints on GC Pulsar Populations}
\title{Multiwavelength Constraints on Pulsar Populations in the Galactic Center}

\author{R.S. Wharton\altaffilmark{1,*}, S. Chatterjee\altaffilmark{1}, J.M. Cordes\altaffilmark{1}, J.S. Deneva\altaffilmark{2}, T.J.W. Lazio\altaffilmark{3}}
\altaffiltext{1}{Department of Astronomy, Cornell University, Ithaca, NY 14853}
\altaffiltext{2}{Arecibo Observatory, HC3 Box 53995, Arecibo, PR 00612}
\altaffiltext{3}{Jet Propulsion Laboratory, California Institute of Technology, M/S 138-308, 4800 Oak Grove Dr., Pasadena, CA 91109}
\altaffiltext{*}{Email: \texttt{rwharton@astro.cornell.edu}}

\begin{abstract}
\noindent The detection of radio pulsars within the central few parsecs 
of the Galaxy would provide a unique probe of the 
gravitational  and magneto-ionic environments 
in the Galactic center (GC) and, if close enough to \sgra, precise tests 
of general relativity in the strong-field regime.  
While it is difficult to find pulsars at radio wavelengths because of 
interstellar scattering, the payoff from detailed timing of pulsars in the GC
warrants a concerted effort.  To motivate pulsar surveys and help define
search parameters for them, we 
constrain the pulsar number and spatial distribution using a wide
range of multiwavelength measurements. 
These include the five known radio pulsars within 15\arcmin\ of \sgra, 
non-detections in high-frequency pulsar surveys of the central parsec, 
radio and gamma-ray measurements of diffuse emission, 
a catalog of radio point sources from an imaging survey, 
infrared observations of massive star populations in the central few parsecs, 
candidate pulsar wind nebulae in the inner 20~pc, 
and estimates of the core-collapse supernova rate based on X-ray measurements.  
We find that under current observational constraints, the inner parsec of the Galaxy 
could harbor as many as ${\sim}10^3$ active radio pulsars that are beamed toward Earth.  
Such a large population would distort the low-frequency measurements of both the intrinsic 
spectrum of \sgra\ and the free--free absorption along the line of sight of \sgra.

\end{abstract}


\maketitle

\section{Introduction}
The discovery of one or more pulsars in the inner parsecs around
Sgr~A*, the massive black hole (MBH) at the center of our
Galaxy, would provide an invaluable tool for studying the innermost
regions of the Galactic center (GC).  Most of the current understanding
of the inner parsec comes from infrared observations of the nuclear
star cluster \citep[for a recent review, see][]{geg10}.  The nuclear
star cluster is centered on \sgra\ and consists of young massive stars
at a projected radius of $r\approx0.5$~pc and a dense collection of
B-stars (the ``S-stars'') within $r\leq0.04$~pc with the closest
orbit passing just $6\times10^{-4}$~pc (${\approx}100$~AU) from \sgra\
\citep{sog02, gdm03}. Two decades of monitoring the orbits of these S-stars
has yielded the mass of the central object to be $M=4\times10^6M_{\odot}$,
unambiguously classifying it as a MBH \citep{gsw08, get09}.

Despite the success of tracking stellar orbits in the infrared, the
sensitivity of this method is ultimately limited by source confusion.
The detection of a radio pulsar at a similar distance with an orbital
period of $P_{orb}\lesssim100$~yr would provide unparalleled tests of
gravity in the strong-field regime.  The timing of such a pulsar
could allow the measurement of the spin or quadrupole moment of the
MBH \citep{pl04, lw97, wk99, lwk12}.  Additionally, a pulsar found anywhere in the inner few parsecs
of the Galaxy would provide a useful probe of the GC environment.
The mere detection of a pulsar would place constraints
on the star formation history (SFH) and measurements of
the dispersion measure and pulse broadening times would provide
information on the electron density distribution of the region.

However, even with the detection of almost 2000 radio pulsars in the Galaxy 
\citep{PSRCAT} and 
several directed searches of the GC, only five pulsars have been found within 15{\arcmin}
of  \sgra\ and the closest of these is $11^{\prime}$ away \citep{dcl09, jkl06, bjl11}. 
While these few objects indicate the existence of a GC pulsar population, 
the perceived dearth of pulsars near Sgr~A* is the result of 
interstellar scattering from turbulent plasma,
which temporally broadens pulses to approximately $2000\nughz^{-4}$~s 
%$6.3\nu^{-4}~\mbox{s}$ 
(where $\nughz$ is the observing frequency in GHz) at the center of the Galaxy \citep{cl02}.  Pulse broadening makes it almost impossible to detect even
long-period pulsars in periodicity searches at commonly used frequencies 
($\nu\sim 1$~GHz).    
To mitigate the deleterious effects of interstellar scattering, 
periodicity searches of the GC have migrated to higher frequencies 
($\nu \sim 10~\mbox{GHz}$).  However, since pulsars have power-law 
spectra of the form $S(\nu)\propto\nu^{\alpha}$ (with $\alpha<0$), 
increasing the observing frequency also decreases the observable flux density.  
To date, high-frequency searches have produced no new detections 
using existing 100~m class telescopes \citep{deneva10, mkfr10}. 

\begin{figure}[t]
\begin{center}
\includegraphics[width=0.5\textwidth]{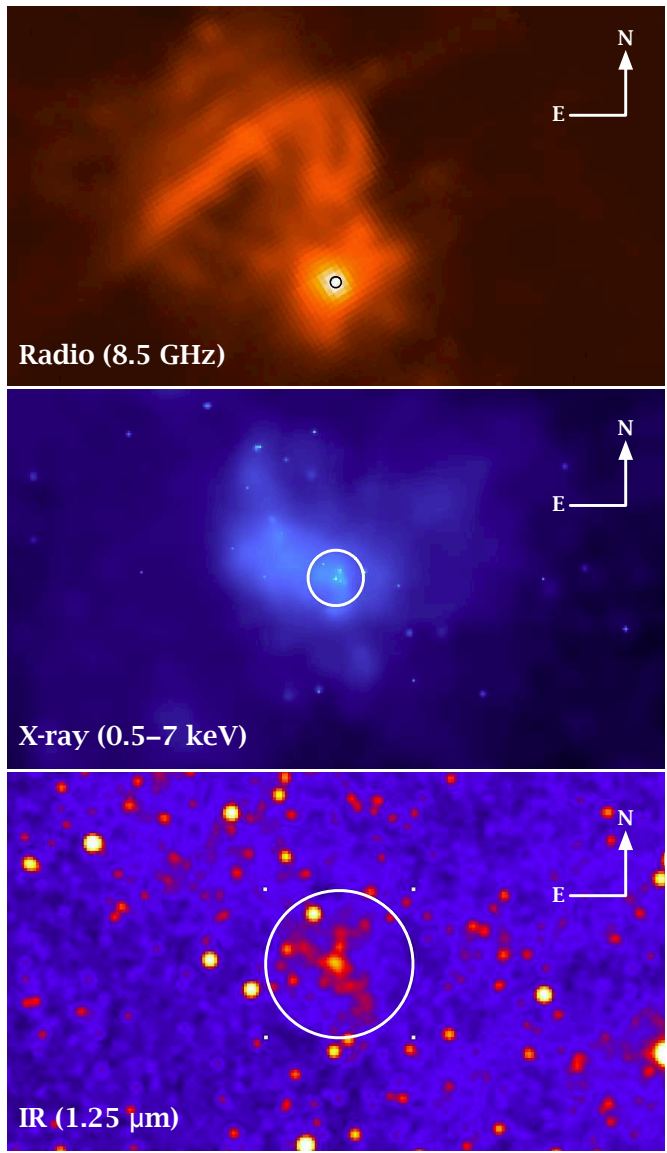}
\caption{Views of the inner GC region at radio, X-ray and infrared wavelengths.  From top to bottom, the GC is shown in radio at 8.5~GHz as observed with the Green Bank Telescope (data courtesy Casey Law), in 0.5--7~keV X-rays as observed with the \emph{Chandra} ACIS-I instrument, and at J band ($1.25~\mu$m) as observed in 2MASS \citep{2MASS}.  In each panel, a circle with radius {25\arcsec} (corresponding to 1~pc at 8.5~kpc) is centered on {\sgra} with J2000 coordinates of ($17\h45\m40\fs0409$, $-29\degr00\arcmin28\farcs118$) given by \citet{rb04}.}
\label{fig:beam}
\end{center}
\end{figure}

Even though the absence of pulsar detections in the central parsecs of the 
GC is well explained by scattering effects,  the existence of a GC pulsar 
population was established by \citet{dcl09} based on the five pulsars on the outskirts of the region that cannot be explained as foreground disk objects.
Since future surveys can benefit from better knowledge of the pulsar populations in the GC, 
we use a suite of multiwavelength observations to set constraints on the number and distribution of pulsars in the inner regions of the Galaxy on ${\sim}100$~pc and ${\sim}1$~pc scales.  An illustration of the structure of the GC on these scales is shown in Figure \ref{fig:beam}.

In this paper, we present observational constraints on the pulsar populations in the GC.  
A brief overview of the conventions and notations used in this paper are presented in Section~\ref{notation}.  
In Section~\ref{radio_surveys}, population limits are set from the detections of pulsars in the inner 15{\arcmin} and from non-detections in the vicinity of {\sgra} on parsec scales.  
In Section~\ref{point_sources1}, a catalog of steep-spectrum radio sources in the inner 150~pc is considered. 
Interferometric measurements of the spectrum of {\sgra} are used in Section~\ref{diffuse_radio} to set upper limits on the pulsar population in the GC on arcsecond scales.  
In Section~\ref{diffuse_gamma}, \emph{Fermi} observations of the diffuse gamma-ray flux of the inner degree of the GC are used to estimate the millisecond pulsar population in the GC.  
In Section~\ref{massive_stars}, infrared observations of young massive stars are used to estimate the number of neutron stars produced in the inner parsec of the Galaxy.  
\emph{Chandra} X-ray observations of pulsar wind nebulae are used to constrain the pulsar population in the inner 20~pc in Section~\ref{PWN_sec}. 
In Section~\ref{supernova}, limits are set on the intrinsic neutron star population in the GC based on the estimated supernova rate.  
Finally, in Section~\ref{discussion}, the estimates are summarized and discussed.

\section{Conventions and Notations}
\label{notation}
We note briefly a few conventions and notations that we adopt in this paper.  The name ``\sgra'' is used to describe both the MBH and the compact radio source at the dynamical center of the Galaxy.  In the event where a distinction must be made (for example, Section~\ref{diffuse_radio}), ``\sgra'' is taken to mean the observed radio source.  All distances from Sgr~A* are given as a projected distance unless explicitly stated otherwise.  The distance from Earth to \sgra\ is taken to be $d = 8.5~\mbox{kpc}$ \citep{gsw08, get09}.  

Pulse broadening times are taken from the NE2001 electron density model of \citet{cl02}.  The NE2001 model accounts for the geometry of the scattering region in the GC, and as such, it is preferable over empirical fits to pulsars in the Galactic disk (e.g. \citealt{bcc04}), which do not consider the particular scattering geometry of the GC.  As discussed in \citet{dcl09}, NE2001 tends to overestimate the scattering times for the five pulsars closest to \sgra\ by factors of ${\sim}10^2-10^3$.  However, these pulsars likely lie along the edges of the GC scattering region where slight (${\sim}0.1$~kpc) changes to the line of sight distance can cause dramatic (factors of ${\sim}10^4$) changes in scattering times with only modest (factor of ${\sim}2$) changes in dispersion measure.  Additionally, we note that the scattering times of pulsars deeper in the GC will be highly constrained by the measured angular scattering of \sgra\ itself.

Pulse broadening times are found at arbitrary observing frequencies by scaling the 1~GHz values given by NE2001 as $\propto\nu^{-4}$ \citep{lr99}.   However, see \citet{lkm01} for potentially significant deviations from this scaling in highly scattered pulsars.

All pulsar population estimates are given as the number of active radio pulsars beamed toward the Earth, where a beaming fraction of $f_b = 0.2$ is assumed for all pulsars.  The fixed beaming fraction of $f_b=0.2$ comes from a simple model in which the magnetic dipole moment is oriented randomly with respect to the rotation axis \citep{ec89}.  A better empirical fit to the data is provided by the period-dependent model of \citet{tm98}, which typically finds $f_b\sim0.1$ for isolated pulsars.  However, since the \citet{tm98} model does not include millisecond pulsars that may have beaming fractions as high as $f_b\approx0.5-0.9$ \citep{kxl98}, we adopt the $f_b=0.2$ value as a population-wide representative value.  As this factor only shows up in our analysis as a multiplicative constant, it is trivial to scale our results to different beaming fractions.

In many of the limits presented below, it will be necessary to utilize a distribution for pulsar pseudo-luminosities ($L=Sd^2$).  We adopt the power-law distribution of \citet{parkesVI2006} with a minimum cutoff in preference to distributions that do not require cutoffs like the log-normal model of \citet{fgk06}.  Both are two parameter models and the power-law distribution provides a much better empirical fit to the observations \citep[see, e.g., Figure 6 of][]{parkesVI2006}.  However, if the log-normal distribution were used instead of the power-law in the calculations below, the total pulsar populations predicted would be larger by factors of 10--100 owing to the much larger fraction of low luminosity objects.  As a result, our adoption of the power-law pseudo-luminosity distribution is a conservative one in the sense that using another distribution would predict a \emph{larger} pulsar population.

Lastly, we note that in some cases our constraints will involve all types of pulsars, while others involve one of two subsets of pulsars:  the ``canonical'' pulsars (CPs) and recycled or ``millisecond'' pulsars (MSPs).  
CPs have periods of $P\sim1~\mbox{s}$, surface magnetic fields 
$B\sim10^{12}~\mbox{G}$ and active radio lifetimes of 
$\tau\sim10^7~\mbox{years}$, 
while MSPs have periods $P\lesssim10~\mbox{ms}$, low surface magnetic fields $B\lesssim10^9~\mbox{G}$ and active radio lifetimes of
$\tau\sim 10^9-10^{10}~\mbox{years}$.

\section{Constraints from Pulsar Surveys of the GC}
\label{radio_surveys}

Several pulsar searches have been conducted in the inner degree of the Galaxy.  To date, only five pulsars have been detected within $15^\prime$ of \sgra, with none closer than $11^\prime$ \citep{PSRCAT}.  
To have any chance of making detections in the inner few arcminutes of the Galaxy, higher observing frequencies must be used to overcome the roughly $2000\nughz^{-4}$~s broadening times caused by scattering.  Deep searches have been attempted at frequencies from 4 to 15~GHz, but have made no detections 
in the inner few parsecs around {\sgra}  
\citep{jkl06,deneva10,mkfr10}. 
Both the pulsar detections in ``low-frequency'' (2--3 GHz) surveys and the absence of detections in high-frequency (4--15 GHz) directed searches can be used to constrain the GC pulsar population.

\subsection{GC Pulsar Detections in Low-Frequency Surveys}
\label{ssec:lowfreq}
The five known pulsars within $15^\prime$ of {\sgra} currently provide the best direct evidence for an intrinsic GC pulsar population.  Of these five pulsars, two were detected at 3.1~GHz with the Parkes radio telescope \citep{jkl06} and three\footnote{One of these pulsars, J1746$-$2850, was independently discovered by \citet{bjl11} in a Parkes 6.5~GHz multibeam survey.} were detected at 2~GHz with the Green Bank Telescope \citep{dcl09}.  In each of these surveys, the expected number of detectable disk pulsars in the field of view is $\ll1$.  Thus, the detections strongly suggest a pulsar population in the GC that is distinct from that of the disk.

In an attempt to constrain the number and spatial distribution of the GC pulsars, \citet{dcl09} simulated the pulsar population to determine what would be consistent with the survey detections.  A simple two-component density model of the form
\begin{equation}
\label{ellipsoidal}
n_{GC}\propto\exp\left(-\frac{h^2}{H_{GC}^2}\right)\exp\left(-\frac{r^2}{R_{GC}^2}\right)
\end{equation}  
was adopted with $H_{GC}=26$~pc fixed to coincide with the scale height of the scattering screen in the NE2001 electron density model of \citet{cl02}.  The density distribution was normalized so that, for a given $R_{GC}$, there were a total of $N_{GC}$ pulsars (not necessarily beamed toward Earth) associated with the population.  

Using Monte Carlo methods, \citet{dcl09} then generated 1000 pulsar populations consistent with Equation~\ref{ellipsoidal} for each $(R_{GC}, N_{GC})$ pair and determined how many of these pulsars would have been detected in their 2.1~GHz survey. Searching over a grid of values and performing a maximum likelihood analysis, \citet{dcl09} found lower bounds of $N_{GC}\gtrsim2000$ and $R_{GC}\gtrsim0.3$~kpc for the parameters.  Thus, this analysis provides additional evidence for an intrinsic pulsar population in the GC.

Although \citet{dcl09} have set a lower bound on $N_{GC}$, this parameter applies to the entire GC population and does not necessarily translate into a lower bound on the number of pulsars within a particular distance from {\sgra}.  For example, a distribution with $N_{GC}=2000$ and $R_{GC}=0.3$~kpc will produce a very different number of pulsars within 100~pc of {\sgra} than will a distribution with $N_{GC}=2000$ and $R_{GC}=0.6$~kpc.  Regardless, the existence of a pulsar population in the GC is firmly established.

We note briefly that the primary reason the \citet{dcl09} analysis does not find upper bounds on the parameters $N_{GC}$ and $R_{GC}$ even after extending the grid to $N_{GC}=10^4$ and $R_{GC}=5$~kpc is that the only constraints come from the detections in the survey region ($r\lesssim50$~pc).  Incorporating survey results from the inner few degrees of the Galaxy would certainly introduce upper bounds to the parameters.  

With this in mind, it is instructive to consider the results at a fixed $R_{GC}$.  From Figure 4 of \citet{dcl09}, we see that a wide range of $N_{GC}$ values with $N_{GC}\gtrsim500$ are equally likely for $R_{GC}=0.1$~kpc.  Thus, a very conservative lower bound on the number of pulsars in the inner 100~pc of the Galaxy that are beamed toward Earth is $N_{psr}\gtrsim100$ (where we have assumed a beaming fraction of 0.2).

\subsection{High-Frequency Pulsar Searches of the Central Parsec}
\label{gbt_search}
Recently, searches of the central few parsecs around \sgra~have been conducted with the Green Bank Telescope (GBT) at 5~GHz and 9~GHz \citep{deneva10} and at 15~GHz \citep{mkfr10}.  
No pulsars were detected in any of these searches.  We follow a similar analysis to that of \citet{mkfr10} to estimate an upper limit to the pulsar population based on the absence of detections.      

\subsubsection{Observations}
\label{sssec:obs}
The 5 and 9~GHz observations were carried out by \citet{deneva10} in 2006.  Since no pulsar candidates were detected, limits on the flux density of periodic signals may be set using the radiometer equation
\be
\label{eq:flux_max}
S_{min,\nu} = \frac{mT_{sys}}{\eta G \sqrt{N_h N_{pol} \Delta\nu T_{obs}}},
\ee
where $T_{sys}$ is the system temperature, $G$ is the telescope gain, $\eta\approx0.8$ is a correction factor that accounts for system imperfections and the digitization of the signal, $N_h=16$ is the maximum number of harmonics summed in a periodicity search, $N_{pol}=2$ is the number of polarization channels summed, $\Delta\nu=800$~MHz is the receiver bandwidth and $T_{obs} = 6.5$~hr is the observation time.  The telescope gain is given by $G=1.85~\mbox{K Jy}^{-1}$ and $G=1.8~\mbox{K Jy}^{-1}$ for observing frequencies of 5 and 9~GHz, respectively.  The value of $m$ is determined by the detection significance threshold set at $m\sigma$.  In this FFT search the threshold was set to $6\sigma$, so $m=6$.

The system temperature of the telescope is given by\footnote{We take $T_{rec}$ to include all non-astronomical contributions to the system temperature from the receiver, spillover effects, and the atmosphere. }  $T_{sys} = T_{rec}+T_{bg}$.  
The receiver temperature of the GBT\footnote{As provided in the GBT Proposer's Guide: \texttt{http://www.gb.nrao.edu/gbtprops/man/GBTpg.pdf}}  
is 18~K at 5~GHz and 27~K at 9~GHz.  The dominant contribution to the background temperature is the bright extended Sgr~A~Complex (comprised of Sgr~A~East and Sgr~A~West), which surrounds \sgra.  
We may set lower bounds on this background using data from a multiwavelength survey by \citet{lyc08}, which imaged the GC at 1.4, 5 and 9~GHz.  
Only lower bounds may be set since a non-trivial iterative scheme was used to subtract out the noise contributions from the atmosphere.    
\citet{lyc08} found the flux density of the Sgr~A Complex to be $85~\mbox{Jy beam}^{-1}$ at 5~GHz and $39~\mbox{Jy beam}^{-1}$ at 9~GHz, which translate to background temperatures of $T_{bg}=157$~K and $T_{bg}=70$~K, respectively.  Thus, the system temperature is $T_{sys}=175$~K at 5~GHz and $T_{sys}=97$~K at 9~GHz.  From Equation~\ref{eq:flux_max}, we find that $S_{min,\nu} = 29~\mu$Jy at 5~GHz and $S_{min,\nu} = 17~\mu$Jy at 9~GHz.

Likewise, \citet{mkfr10} conducted a search for pulsars at 14.8 and 14.4~GHz using the GBT in 2006 and 2008.  The \citet{mkfr10} search used a $10\sigma$ detection threshold and combined observations to get an effective observation time of $T_{obs}\approx9.75$~hr.  The system temperature was determined to be $T_{sys}\approx35$~K by firing a noise diode on a calibrator source.  No pulsars were detected and a $10\sigma$ detection threshold flux density of $S_{min,\nu}\approx10~\mu$Jy is set from the 14.4~GHz measurements.

\subsubsection{Upper Limits on Observable Pulsar Population}    
\label{sssec:hf_ul}
Assuming that a given pulsar in the GC will be detected with some probability $p_d$, binomial statistics can be used to find the maximum number of pulsars consistent with zero detections.  
The simplest way to determine $p_d$ is to set it equal to the fraction of pulsars bright enough to be seen in each survey when placed at \sgra.  This fraction can be estimated from the 1.4~GHz pseudo-luminosity function, which is given by 
$dN/d \log{L} \propto L^{-\beta}$
taken over a range of pseudo-luminosities from
$L_{min}=0.1~\mbox{mJy}~\mbox{kpc}^2$ to $L_{max} = 10^4~\mbox{mJy}~\mbox{kpc}^2$.
We adopt a value of $\beta=-0.7$ as an average of the two fits found by \citet{parkesVI2006} of over 1000 pulsars observed in the Parkes Multibeam Survey. 
The detection probability is then given by $p_d=f_L(L>L_{det})$, where
\be
\label{eq:ldetect}
L_{det} = S_{min,\nu}\left(\frac{1.4~\mbox{GHz}}{\nu}\right)^{-1.7}d^2_{gc}
\ee 
and $S_{min,\nu}$ is the minimum detectable flux density of the search.
Using the $S_{min,\nu}$ values found by each search (see Table~\ref{tab:hf_vals}) we find that $p_d=(0.027, 0.020, 0.015)$ at $\nu=(5,9,15)$~GHz.

Given the above detection probabilities and the lack of any detections in the surveys, the upper limits to the number of pulsars (at 99\% confidence level) are found to be $N=(170, 230, 299)$ for the 5, 9, and 15~GHz observations, respectively.  However, since a pulsar with a spin period less than the pulse broadening time would have a greatly reduced chance of being detected, the calculated upper limits are for pulsars with $P\gtrsim\tau_{sc}$.  Using the 1~GHz scattering time from the NE2001 model of \citet{cl02} and scaling ($\propto\nu^{-4}$) to the appropriate frequency, the scatter broadening times are found to be $\tau_{sc}=(4.2, 0.44, 0.05)$~s at observing frequencies of $\nu=(5,9,15)$~GHz.

%\ctable[
%	caption = Pulsar Upper Limits and Survey Parameters,
%	label     = tab:hf_vals,
%	pos	    = h,
%	center
%]{ccccccccc}{\tnote[a]{\citet{deneva10}} \tnote[b]{\citet{mkfr10}}}
%{\toprule
%\toprule
%$\nu$ & $S_{min}$ & $\tau_{sc}$ &  $f_L(L>L_{det})$ & $N(P\gtrsim\tau_{sc})$ & $f_{P}(P>\tau_{sc})$ & %$N/f_P$ & $r$ & Ref\\
%(GHz) & $(\mu$Jy) & (s) & & &  &  & (pc) & \\
%(1) & (2) & (3) & (4) & (5) & (6) & (7) & (8) & (9)\\
%\cmidrule{1-9}
%4.85 & 29 & 4.2 & 0.027 & 170 & 0.06 & 2830 & 3.0 & (1)\tmark[a]\\
%8.50 & 17 & 0.44 & 0.020 & 230 & 0.43 & 535 & 1.5 & (1)\tmark[a]\\
%14.4 & 10 & 0.05 & 0.015 & 299 & 0.63 & 475 & 1.0 & (2)\tmark[b]\\
%\bottomrule
%}

\begin{table*}[ht]
\begin{center}
\begin{threeparttable}
\caption{Pulsar Upper Limits of High-Frequency Surveys of the Central Parsecs} 
\label{tab:hf_vals}
\begin{tabular}{ccccccccc}
\toprule
\toprule
$\nu$ & $S_{min}$ & $\tau_{sc}$ &  $f_L(L>L_{det})$ & $N(P\gtrsim\tau_{sc})$ & $f_{P}(P>\tau_{sc})$ & $N/f_P$ & $r$ & Ref\\
(GHz) & $(\mu$Jy) & (s) & & &  &  & (pc) & \\
(1) & (2) & (3) & (4) & (5) & (6) & (7) & (8) & (9)\\
\cmidrule{1-9}
4.85 & 29 & 4.2 & 0.027 & $<170$ & 0.06 & $<2830$ & 3.0 & (1)\\
6.6 & 550\tnote{$\dagger$} & 1.28 & 0.002 & $<2220$ & 0.17 & $<13000$ & 4.0 & (2) \\
8.4 & 200 & 0.46 & 0.003 & 	$<1380$ & 0.43 & $<3200$ & 3.0 & (3) \\
8.50 & 17 & 0.44 & 0.020 & $<230$ & 0.43 & $<535$ & 1.5 & (1)\\
14.4 & 10 & 0.05 & 0.015 & $<299$& 0.63 & $<475$ & 1.0 & (4)\\
\bottomrule
\end{tabular}
\begin{tablenotes}
\footnotesize
\item[] {\bf Notes.} Columns 1 and 2 give the center frequency and minimum detection threshold, respectively, of a given survey.  Column 3 gives the scattering time at the survey center frequency of a pulse originating from the GC as calculated with the NE2001 model \citep{cl02}.  Column 4 gives the fraction of pulsars bright enough to be detected at the GC with the survey sensitivity (see Equation~\ref{eq:ldetect}).  Column 5 gives 99\% upper limit values to the number of pulsars that could potentially have been seen by the survey (that is, have $P\gtrsim\tau_{sc}$).  Column 6 gives the fraction of pulsars in the ATNF catalog \citep{PSRCAT} that have periods longer than the scattering time and Column 7 gives an estimated upper bound for the total number of pulsars in the GC.  Column 8 gives the projected radial distance of the beamsize at the distance of {\sgra} and Column 9 gives the reference from which the survey was taken.
\item[] ${}^\dagger$The value provided for $S_{min}$ here is 9 times that given in Section~2.2 of \citet{bjl11}, which we believe to be a calculation error.
\item[] {\bf References.} (1) \citealt{deneva10}; (2) \citealt{bjl11}; (3) \citealt{jkl06}; (4) \citealt{mkfr10}
\end{tablenotes}
\end{threeparttable}
\end{center}
\end{table*}

\subsubsection{Implications for Total GC Pulsar Population}

The upper limits of Section~\ref{sssec:hf_ul} are only valid for pulsars within certain period ranges.  To make an estimate of the total number of pulsars, knowledge of the underlying pulsar period distribution is needed.  Though the period distribution of pulsars in the inner parsecs of the GC is entirely unknown, a reasonable approximation would be to assume the same distribution as the local pulsar population (to reduce observational biases).  From the ATNF 
pulsar catalog\footnote{\texttt{http://www.atnf.csiro.au/research/pulsar/psrcat/}}
\citep{PSRCAT}, we see that there are 88 pulsars within 1~kpc of Earth.  Of these, 5 have $P>4.2$~s, 38 have $P>0.44$~s and 55 have $P>50$~ms.  This gives $f_P(P>4.2~\mbox{s}) = 0.06$, $f_P(P>0.44~\mbox{s})=0.43$ and $f_P(P>50~\mbox{ms})=0.63$ for the fractions of pulsars with periods greater than the scatter broadening times.

Assuming these values are representative of the GC population, we can estimate upper bounds on the total number of pulsars (regardless of period) to be $N_{max}=N/f_P$.  Applying these corrections to the estimates of Section~\ref{sssec:hf_ul} gives $N_{max} = (2830, 535, 475)$ for observing frequencies of $\nu=(5,9,15)$~GHz.

The half-power beam width of the GBT is $\theta(\nu)\approx150^{\arcs}(\nu/5~\mbox{GHz})^{-1}$.  At \sgra, the projected radii of these beams are $r\approx(3.0, 1.5, 1.0)$~pc at 5, 9 and 15~GHz, respectively.  Thus, we estimate that there are as many as $N_{max}<2830$ pulsars beamed toward Earth within $r\approx3.0$~pc of \sgra, $N_{max}<535$ within $r\approx1.5$~pc and $N_{max}<475$ within $r\approx1$~pc.

The results of this section are summarized in Table \ref{tab:hf_vals}.  In addition to the surveys of \citet{deneva10} and \citet{mkfr10}, we also include for reference the less sensitive GC pointings from surveys by \citet{jkl06} and \citet{bjl11}.

\subsubsection{Caveats}
A number of assumptions are made in the upper limit estimates of the previous sections, so it is important to consider what happens if the assumptions fail.  The first assumption is that the probability of detecting a pulsar (with period large enough not to be smeared out by interstellar scattering) is equal to the fraction of pulsars with 1.4~GHz pseudo-luminosities large enough to be detected at the GC.  Since the detection probability only considers the best-case sensitivity of the telescope and ignores any effects of radio-frequency interference at the telescope end or intermittency at the pulsar end, it is likely to be an overestimate.  An overestimate of the detection probability would result in an underestimate of the upper bound on the number of pulsars.

Another assumption is that the 1.4~GHz pseudo-luminosity distribution of pulsars in the GC is the same as those in the Galactic field.  However, if the GC region contains a larger number of low-luminosity MSPs, then the detection probability will again be overestimated and the upper limit underestimated.  One could also imagine the case where enough bright young pulsars exist in the GC as a result of recent star formation to skew the pseudo-luminosity distribution to higher luminosities.  In that case, the detection probability would be an underestimate and the upper limits an overestimate.

We have also assumed that the upper limits are for pulsars with $P>\tau_{sc}$. However, as the scattering time gets to be a significant fraction of the pulsar period, it will start to smear out the signal and reduce the number of detectable harmonics. Since this is a gradual process, it will likely make some pulsars undetectable even with $P>\tau_{sc}$ . As a result, we would be overestimating the minimal detectable period, which would cause the upper bound to be an underestimate. 

Finally, we have assumed that the period distribution of the pulsars within 1~kpc of Earth is representative of the GC population. This certainly does not have to be the case, as the star formation histories of the Galactic field and GC are likely to be different. For example, the GC could potentially have a much higher concentration of young pulsars and old MSPs as compared to the Galactic disk.  The increased stellar encounter rate in the GC could favor MSP production and any recent starburst would favor young pulsars.  In both cases, the periods would be biased low. Thus, the period distribution would be skewed lower than assumed and the fraction of pulsars with periods greater than a certain value will be overestimated. This will result in the upper bound being underestimated.

Since most of the assumptions made tend to decrease the upper limits, our estimates are best interpreted as the most restrictive upper bounds to the pulsar population in the inner few parsecs of the GC.  

\section{Radio Point Sources}
\label{point_sources1}
Motivated by the study of GC pulsar search methods by \citet{cl97}, \citet{lc08} performed a VLA survey of compact radio sources in the inner degree of the Galaxy.  Though pulsars cannot be identified by their pulsed emission in an imaging survey, promising pulsar candidates may be found by looking for steep-spectrum sources with angular diameters consistent with the angular broadening of point sources caused by scattering at locations near {\sgra} (${\approx}1\arcsec$~at 1~GHz).  Of the 170 compact radio sources cataloged, \citet{lc08} estimate that the number of pulsars included is of order ${\sim}10$.  Based on this survey, upper limits to the pulsar population within $1^\circ$ (${\approx}150$~pc) of {\sgra} may be estimated.  

\subsection{Observations}
The survey was conducted at observing frequencies of 1.4 and 5~GHz with the VLA in the A~configuration.  A total of 13 fields arranged in a hexagonal grid covered the region of the GC out to roughly $1^\circ$  (150~pc) from {\sgra} (the half-power radius of the VLA primary beam is $15\arcmin$ at 1.4~GHz).  The typical resolution for the survey was a synthesized beam size of $2\farcs4\times1\farcs3$.

Sources were identified using a method similar to that of \citet{lc98}.  Essentially, a histogram of intensities was constructed from the image of the primary beam for each field.  If the field just contained noise, the intensity histogram would be a Gaussian with a mean of zero and a standard deviation equal to the thermal noise of 0.05~mJy per synthesized beam.  Sources could then be determined by looking for deviations from this noise-only histogram.  In practice, the histogram was found to have larger tails than a Gaussian, with zero mean and a standard deviation of ${\approx}0.5~\mbox{mJy beam}^{-1}$.  Since the resolution is comparable to the scattering size of a point source at the distance of {\sgra}, a $10\sigma$ detection threshold of $S_{det}\approx5$~mJy was adopted for the survey.

\subsection{Pulsar Population Estimate}
Given that $N_{obs}\sim10$ pulsars were likely observed, the total pulsar population in the survey region can be estimated as $N_{psr} \sim N_{obs}/f_L$, where $f_L$ is the fraction of pulsars luminous enough to be detected at the distance of {\sgra}.  Taking the survey detection threshold to be $S_{det}=5$~mJy at 1.4~GHz, a pulsar must have a 1.4~GHz pseudo-luminosity of at least $L_{det}=360~\mbox{mJy kpc}^2$ to be detected at the distance of {\sgra}.    
The fraction of pulsars with $L>L_{det}$ can be determined from the 1.4~GHz pulsar luminosity function, which has the form 
$dN/d \log{L} \propto L^{-\beta}$.  
The range of pulsar pseudo-luminosities is taken from 
$L_{min} = 0.1~\mbox{mJy}~\mbox{kpc}^2$ to  $L_{max} = 10^4~\mbox{mJy}~\mbox{kpc}^2$ and the exponent in the distribution function is taken to be $\beta = 0.7$ \citep{parkesVI2006}.  From this distribution, the fraction of pulsars luminous enough to be detected is $f_L=3\times10^{-3}$.  For $N_{obs}\sim10$ pulsars detected in the survey, we expect a total population of $N_{psr}\sim3000$ pulsars within $1^\circ$ ($150$~pc) of {\sgra}.

Though $N_{psr}\sim3000$ is the nominal population estimate from the survey, a broader range results if the assumptions do not exactly hold.  For instance, \citet{lc08} estimate that $N_{obs}\sim10$ of the unidentified steep-spectrum point sources will ultimately turn out to be radio pulsars.  However, this number could range from zero to about 30.  If one takes $N_{obs}=30$, repeating the above analysis gives a pulsar population of $N_{psr}\sim10^4$.  Additionally, one may consider the case in which no pulsars were detected.  Despite a follow-up observation of 15 of the pulsar candidates in this survey by \citet{deneva10} with the GBT, none of the candidates have to date been confirmed.  Assuming zero detections, an analysis similar to that in Section~\ref{gbt_search} gives an upper limit to the pulsar population of $N_{psr}\leq1500$ at a 99\% confidence level.  These two extremes illustrate that although the survey allows for an estimate of the pulsar population in the inner degree of about 3000, the actual number could be below 1500 or as high as $10^4$.  As a result, we take $N_{psr}\lesssim10^4$ as a conservative upper bound.

Finally, we note that although the survey covers the region within $1^\circ$ ($150$~pc) of {\sgra}, there will be reduced sensitivity in the field centered on {\sgra}.  The reduced sensitivity is the result of increased background temperatures and greater sidelobes from the extended structure of the inner GC.  In addition, since the scatter-broadening of point sources in the vicinity of {\sgra} (${\approx}1\arcsec$) is comparable to the resolution of the survey, source confusion may become important in the innermost regions of the GC.  Therefore, this survey would be largely insensitive to a fairly compact population of pulsars in the inner tens of arcseconds around {\sgra}.

\section{Radio Spectrum of Sgr A*}
\label{diffuse_radio}
In this section, limits are placed on the maximum allowable number of
pulsars in the inner parsecs of the GC based on radio interferometer
observations of {\sgra} on arcsecond scales ($1\arcsec\approx0.04$~pc
at 8.5~kpc).  Due to the finite resolution of interferometers
and the broadening of angular diameters as a result of the interstellar
scattering of radio waves, the {\sgra} radio source is actually	
extended (${\approx}1\arcsec$ at 1~GHz).  Flux measurements
of {\sgra} will therefore include a contribution from a collection of
pulsars, if such a population exists.  Although these pulsars will
be unresolved, upper limits on the total population may be set based
on the total flux density of {\sgra} in a manner analogous
to similar constraints placed on pulsars in globular clusters \citep{fg90}.

We consider a model in which the observed flux density of {\sgra} is
actually the combination of two components.  The first component is
that due to radio emission from the immediate environment
of the MBH itself, which we assume
is described accurately by high frequency observations where the pulsar component is negligible.
The second component is that due to the population of
pulsars near the MBH.  This pulsar component becomes
important at lower frequencies both because radio pulsars
typically have steep spectra 
and the angular resolution of radio telescopes scales with frequency such
that a larger region around the MBH is sampled at
lower frequencies.  By requiring that this model flux be consistent with
existing observations, constraints may be set on the maximum number of
pulsars allowed in the inner parsec of the Galaxy.

\subsection{Observations}
Two different measurements of the spectrum of Sgr~A* over a wide 
range of frequencies are considered \citep{an05, falcke98}.  \citet{an05} conducted simultaneous measurements of Sgr~A* from 300~MHz to 43~GHz using the VLA (A-configuration) and the GMRT.  \citet{falcke98} made simultaneous measurements of Sgr~A* using the VLA (A-configuration), the Berkeley-Illinois-Maryland Array, the Nobeyama 45~m telescope, and the Institut de Radioastronomie Millimetrique (IRAM) 30~m telescope from 1.4~GHz to 235~GHz.  Both groups observed a broken power-law spectrum with a break around 10~GHz.  Since the power-law spectrum of pulsars decreases with increasing frequency, a population of pulsars 
contributes significantly only at lower frequencies.  As a result, we consider
the spectrum of {\sgra} only below the break frequency  at 10~GHz.

\subsection{Spectral Model}
\label{ss_model}
We model the measured flux density of the compact radio source \sgra\ as the sum 
of contributions from a collection of pulsars and a point source associated with the MBH attenuated by free-free absorption according to
\begin{eqnarray}
\Ssgr(\nu) &=&\left[ \Sbhvo \left(\frac{\nu}{\nu_0}\right)^{\abh} +N(\nu)\Spsrvo \left(\frac{\nu}{\nu_0}\right)^{\apsr}   \right] \nonumber \\
&& \times\exp( -\nu_{f}^2/\nu^2).
\label{eq:spectrum}
\end{eqnarray}

The emission from the immediate vicinity of the MBH, $\Sbhvo$,
is taken to be a point source with a power-law spectrum 
with spectral index $\abh$. 
$N(\nu)$ is the number of pulsars contained in the
solid angle of the effective point-spread function or beam size, 
which depends strongly on frequency (see Section~\ref{ssec:ang_res}). 
We use a simplified scaling for the free-free absorption that ignores the frequency
dependence of the Gaunt factor.
The free-free absorption factor has a turnover frequency 
$\nu_f$ \citep{an05}.  The flux density per pulsar, $\Spsrvo$, is taken to 
be the mean of the 1.4~GHz pulsar pseudo-luminosity distribution given by $dN/d \log{L} 
\propto L^{-0.7}$ \citep{parkesVI2006} with a lower cutoff of $L_{min} = 0.1 \mbox{ mJy kpc}^2$ and an upper cutoff of $L_{max} = 10^4 \mbox{ mJy kpc}^2$.
The mean observed 
flux density at 1.4~GHz is $\Spsronefour = 99~\mu\mbox{Jy}$.  
Since pulsar radio flux scales 
with frequency as a simple power-law, we can scale our flux density as 
$S_{\nu} \propto \nu^{\apsr}$.  We fix 
$\apsr = -1.7$ as a nominal value for the pulsar spectral index 
\citep{mkk00, lyl95}, but also consider values of $-1.0$ and $-2.5$ to test any major spectral index
dependence.

\subsection{Effective Angular Resolution}
\label{ssec:ang_res}
The number of pulsars included in a flux measurement of \sgra\ can be written
as an integral of the number of pulsars per unit solid angle,
\be
N(\nu) = \int  d\Omega \frac{dn_p}{d\Omega}. 
\ee
The integral is over the effective solid angle 
$\Omegaeff(\nu)= (\pi/4)\theta_{\rm eff}^2(\nu)$, where 
\be
\theta_{\rm eff}(\nu)
	=  \left[\thetab^2(\nu) + \thetasc^2(\nu) \right]^{1/2}
	=  \left(\thetabz^2\nu^{-2} + \thetascz^2\nu^{-4} \right)^{1/2}
\ee
is the effective resolution with the subscript ``0'' representing values at 1~GHz and the frequencies are
in GHz units. The effective resolution is the quadrature sum of the resolution of the interferometer and
the angular extent of \sgra\ caused by scattering.  
The scaling for the synthesized array beam $\thetab$ $\left(\propto\nu^{-1}\right)$ assumes a 
fixed array configuration and the scattering diameter $\thetasc$ $\left(\propto\nu^{-2}\right)$ scales
in conformance to
measurements of \sgra\ and OH/IR masers \citep[e.g.,][]{fcc94}.  
Our treatment assumes that scattered images are circular whereas in fact some
are elliptical, but given that we are making order of magnitude estimates of
pulsar numbers, the differences are not important.

The angular diameter of \sgra\ is dominated by interstellar
scattering at low frequencies, with an observed major axis of  
$\thetascz = 1\farcs2$ \citep{bgf06}.
For the VLA in the A configuration and ignoring any effects of foreshortening, the half-power beam width is $\thetabz = 1\farcs95$ \citep{bridle89}.

The effective resolution of the VLA observations is then
\begin{equation}
\theta_{\rm eff} (\nu) = 
1\farcs2\, \nu^{-2}\left(1 + 2.6~\nu^{2}\right)^{1/2}.    \end{equation}
The two contributions are equal
at $\nu = 0.6$~GHz, so at frequencies lower than this the resolution
is completely scattering dominated and   
the resolution solid angle scales steeply with frequency as $\nu^{-4}$. 

Additionally, a single data point measured with the GMRT will be considered in our analysis, so a similar effective resolution must be constructed for this telescope.  \citet{royrao04} measure the resolution to be $11\farcs4\times7\farcs6$ at 620~MHz.  Converting this ellipse to a circle of equal area and scaling to 1~GHz gives $\thetabz = 5\farcs77$ for the GMRT.  Combining this with scattering as above, gives 
\begin{equation}
\theta_{\rm eff} (\nu) = 
1\farcs2\, \nu^{-2}\left(1 + 23.1~\nu^{2}\right)^{1/2}.    \end{equation}
For the GMRT, the two components of the effective resolution are equal at $\nu=0.2$~GHz.
  
\subsection{Candidate Pulsar Distributions}
\label{ssec:cand_dist}

We choose three physically-motivated distributions as model
pulsar populations.  The distributions are illustrated below in Figure \ref{fig:densities}.

The first (Model A) assumes a constant number of pulsars per unit solid angle,
$dn_p/d\Omega = $~constant,  
so the number of pulsars scales as 
%$N(\theta) \sim {\theta}^{2}$ for angular distance $\theta$ from Sgr~A*.  
\be
N_A(\nu) = N_1 \left[\frac{\Omegaeff(\nu)}{\Omega_1}\right],
\ee
where $\Omega_1$ is the solid angle enclosing $N_1$ pulsars.  In the fitting below, $\Omega_1$ is set so that $N_1$ gives the number of pulsars in the inner parsec.
Referring to Equation~\ref{eq:spectrum}, it may be seen that when scattering dominates the effective 
resolution, the contribution
to the unabsorbed spectrum from pulsars 
increases very rapidly as $N_A(\nu) \nu^{\alpha_{\rm psr}} \propto \nu^{-5.7}$. 
Free--free absorption attenuates much of the flux, thus allowing a
significant pulsar population to remain hidden in spectral measurements. 

In Model B, we assume that $dn_p/d\Omega \propto \Omega^{-0.7}$, 
corresponding to the surface density scaling 
observed for Wolf-Rayet and O-star populations in the 
inner parsec \citep{geg10}.  This yields
\be
N_B(\nu) = N_1 \left[\frac{\Omegaeff^{0.3}(\nu)}{\Omega_1^{0.3}} \right]. \ee

However, the observed populations of Wolf-Rayet and O-stars have an inner cutoff at $\theta\approx1^{\prime\prime}$ \citep{bartko10}.  This core may affect the pulsar population in many ways, but we shall just consider two here.  In both distributions, the pulsar surface density goes as $dn_p/d\Omega \propto \Omega^{-0.7}$ outside the inner cutoff, as before.  Inside the cutoff, one of the distributions (call it Model B-1) has $dn_p/d\Omega = \mbox{const}$ and the other (call it Model B-2) has $dn_p/d\Omega = 0$.  These models give
\be
  N_{B1}(\nu) = \left\{ 
  \begin{array}{l l}
    N_1 \left[\frac{\displaystyle\Omegaeff^{0.3}(\nu) -0.7\Omega_0^{0.3}}
	{\displaystyle\Omega_1^{0.3}-0.7\Omega_0^{0.3}} \right], & \quad \Omegaeff\geq\Omega_0\\
	& \\
     N_1 \left[\frac{\displaystyle0.3\Omega^{0.7}_0\Omega}
	{\displaystyle\Omega_1^{0.3}-0.7\Omega_0^{0.3}} \right]& \quad \Omegaeff<\Omega_0\\
  \end{array} \right.
\ee 
and
\be
  N_{B2}(\nu) = \left\{ 
  \begin{array}{l l}
    N_1 \left[\frac{\displaystyle\Omegaeff^{0.3}(\nu) -\Omega_0^{0.3}}
	{\displaystyle\Omega_1^{0.3}-\Omega_0^{0.3}} \right], & \quad \Omegaeff\geq\Omega_0\\
    0 & \quad \Omegaeff<\Omega_0\\
  \end{array} \right.
\ee 

pulsars enclosed within $\Omegaeff$.

In Model C, we consider a compact population of pulsars contained in a solid
angle much smaller than any resolution solid angle so that
$dn_p/d\Omega$ is effectively a delta function. 
A compact distribution close to \sgra\ could conceivably arise as 
a product of dynamical friction \citep{morris93, meg00}.
Here we simply have
\be
N_C(\nu) = N_1. 
\ee

In addition to the above three, one may consider other models for the pulsar distribution.  For example, the pulsars could be arranged in a central core with a diffuse halo.  However, most of these other distributions can be made as combinations of those we consider.  As a result, we do not expect the final answers to change by more than an order of magnitude.

\begin{figure}[h!]
\begin{center}
\includegraphics[width=0.5\textwidth]{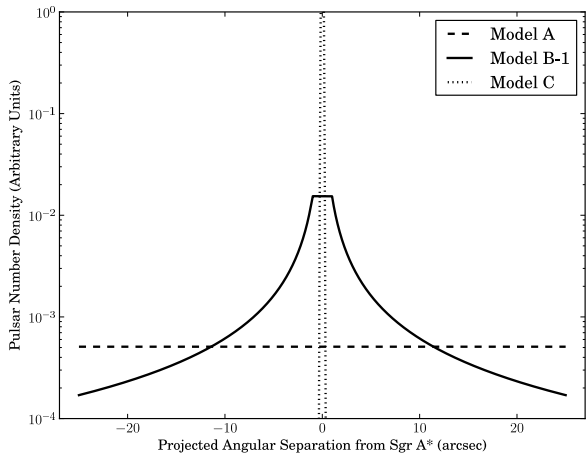}
\caption{Pulsar number density as a function of angular separation from \sgra~for each of the three distribution models.  The distributions have been normalized in this figure so that each model produces the same number of pulsars enclosed within the inner parsec ($\theta_r \approx 25^{\prime\prime}$).  See Section~\ref{ssec:cand_dist} for a description of each model.}
\label{fig:densities}
\end{center}
\end{figure}

\subsection{Model Fitting}
To test our distributions, we calculated the $\chi^2$ values of the model given by Equation~\ref{eq:spectrum} with each of the three distributions against the VLA and GMRT data from \citet{an05} and \citet{falcke98}.  
We fix the pulsar spectral index at $\apsr = -1.7$ and the flux density per pulsar at $\Spsrvo = 99~\mu \mbox{Jy}$ at 1.4~GHz, as described in Section~\ref{ss_model}.  
The flux from the point source associated with the MBH is normalized such that the total measured flux of \sgra\ exactly matches the data at 8.45~GHz.  
The spectral index of the MBH point source ($\abh$), the number of pulsars ($N_1$) and the free-free cutoff frequency ($\nu_f$) are allowed to vary.  
The number of pulsars, $N_1$, is taken within an angular distance of $\theta = 25\arcsec$, which corresponds to a projected radial distance of $r\approx1$~pc from \sgra.  
The allowed ranges for each parameter were chosen to be consistent with current measurements and are presented below in Table~\ref{params}.  
For reference, the best fit values for the model with no pulsars present are $\abh=0.15$ and $\nu_f=0.26$~GHz.    

For each grid point in the three-dimensional parameter space, we 
calculate $\chi^2$ between the measured flux and our model and evaluate the likelihood function assuming independent Gaussian
statistics for measurement errors, 

\begin{eqnarray}
\mathcal{L}(N_1, \abh, \nu_f) &=& \prod_{i=1}^{N_p} \left({2\pi\sigma_i^2}\right)^{-\frac{1}{2}} \exp \left\{ -\frac{\left[\Ssgr(\nu_i) - S_{obs}(\nu_i)\right]^2}{2\sigma_i^2}  \right\} \nonumber \\
&\propto& \exp \left( -\frac{1}{2} \chi^2 \right).
\end{eqnarray}

\begin{figure}[h!]
	\centering
	\includegraphics[width=0.5\textwidth]{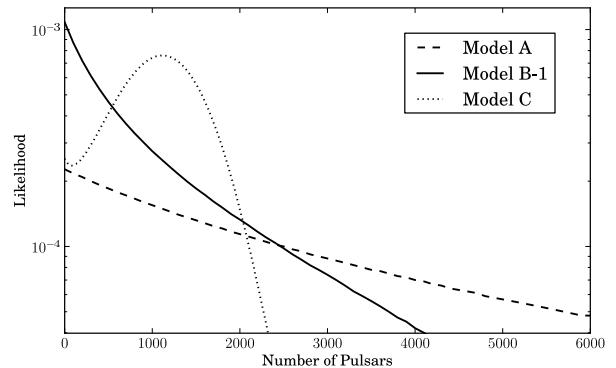}
	\caption{Likelihood functions for three pulsar distribution models (see Section~\ref{ssec:cand_dist} for a description of each model).  The likelihoods have been marginalized over the parameters $\abh$ and $\nu_{f}$ and normalized so that
$\int \mathcal{L}(N)=1$.  
In each case, the pulsar spectral index was taken to be $\apsr=-1.7$.}
	\label{fig:likelihoods}
\end{figure}

The likelihood function is then marginalized over $\abh$ and $\nu_f$ to get a distribution for $N_1$. The marginalized likelihood functions for the number of pulsars within 1 pc of Sgr~A* are plotted in Figure~\ref{fig:likelihoods}.

\resizebox{0.5\textwidth}{!}{
\begin{threeparttable}
\caption{Searched Parameters for Each Pulsar Distribution Model} 
\label{params}
\centering
\begin{tabular}{cccc}
\toprule
\toprule
& \multicolumn{3}{c}{Parameter Ranges\tmark} \\
\cmidrule(l){2-4}
Model & $N_1$ & $\abh$ & $\nu_f~\mbox{(GHz)}$\\
\midrule
A & [ 0, 50000, 100 ] & [ 0.05, 0.40, 0.01 ] & [ 0.05, 1.00, 0.01 ] \\
B & [ 0, 50000, 100] & [ 0.05, 0.40, 0.01 ] & [ 0.05, 1.00, 0.01 ] \\
C & [ 0, 5000, 10 ] & [ 0.05, 0.40, 0.01 ] & [ 0.05, 1.00, 0.01 ]\\
\bottomrule
\end{tabular}
\begin{tablenotes}
\footnotesize
\item[a] Data given as [ min, max, step size ]
\end{tablenotes}
\end{threeparttable}}

\subsection{Results}
\begin{figure}[h!]
	\centering
	\includegraphics[width=0.5\textwidth]{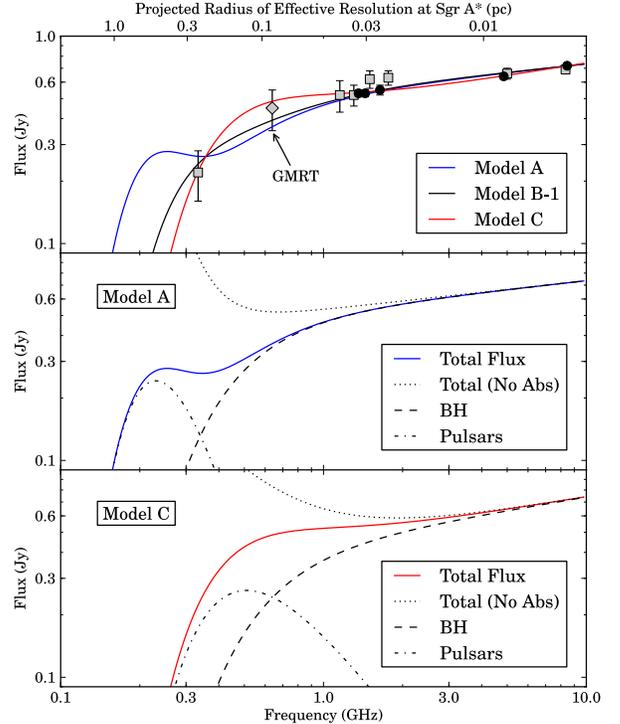}	 						
	\caption{\emph{Top:} Fit curves for model pulsar distributions to the VLA (squares) and GMRT (diamond) data from \citet{an05} and VLA (filled circles) data from \citet{falcke98}.  Since the best fit for Models A and B-1 indicate that the most probable number of pulsars is zero, we have instead plotted the best fit assuming the 68\% confidence upper limit number of pulsars are present. The upper axis gives the projected radial distance of the effective resolution beam at the distance of \sgra~using the VLA (for details, see Section~\ref{ssec:ang_res}).  Note that although the fits were made using both the VLA and GMRT data, the curves above only apply to the VLA data points (squares and filled circles).  \emph{Middle and Bottom:} Components to the observed flux of \sgra\ from the MBH point source and surrounding pulsars for Models A and C.  Note how the pulsar component in Model A rises quickly with decreasing frequency as a result of both the pulsar spectrum and the increasing number of pulsars in the beam.  The total unabsorbed flux is also shown for comparison.}
	\label{fig:flux_plot}
\end{figure}

The best-fit parameters for the maximum likelihood (ML) number of pulsars within the inner parsec of the GC for each model is provided in Table \ref{tab:model_params}
and the resulting spectra are plotted in Figure \ref{fig:flux_plot}.  
The number quoted in Table \ref{tab:model_params} is the ML number of pulsars as determined from the likelihood distribution and the uncertainties denote the most compact 68\% confidence interval around the ML value.  If the ML value for a distribution is zero, then the upper limits are given at the 68\% confidence level.  In addition to the fiducial pulsar spectral index of $\apsr = -1.7$, we include spectral indices of $-1.0$ and $-2.5$.  
For comparison, the best fit parameters for the model with the number of pulsars fixed at zero is also included in Table \ref{tab:model_params} and odds ratios are calculated against this ``null'' model.

It is interesting to note that although Models A and B give only upper limits to the pulsar population (of ${\lesssim}10^3-10^4$), Model C provides a non-zero maximum likelihood value of ${\sim}10^3$.  Additionally, Model C provides a better fit than the model with no pulsars at all (the ``null'' model in Table~\ref{tab:model_params}).  

In all of the models considered, an increase in the maximum number of pulsars is accompanied by an increase in the free--free turnover frequency.  The increased free-free absorption is required to mask the bright low-frequency tail of a large pulsar population.  Current estimates of free--free absorption in the region near \sgra\ give turnover frequencies around 330~MHz \citep{pae89}.  Estimates of the free--free absorption of the flux from \sgra, however, assume a power-law flux density for \sgra\ and would likely underestimate the absorption if a pulsar population were present.   As a result, any independent measurement of the free-free absorption along the line of sight of {\sgra} that does not assume a spectrum of the \sgra\ source could place an important constraint on the pulsar population in the inner parsecs of the Galaxy. 
  
From current radio measurements of the inner parsec of the GC, total pulsar populations (that is, both CPs and MSPs) of up to ${\sim}10^3$ are consistent with observations, regardless of the underlying spatial distribution.

%%\begin{center}
\resizebox{0.5\textwidth}{!}{
\begin{threeparttable}
\caption{Number of Pulsars within 1~pc for Given Model and Spectral Index} 
\label{tab:model_params}
\centering
\begin{tabular}{cccccccc}
\toprule
\toprule
&  & \multicolumn{3}{c}{Model Parameters\tmark[a]} & & &\\
\cmidrule(l){3-5}
Model & $\apsr$ & $N_1 (\times10^3)$ & $\abh$ & $\nu_f$~(GHz) & $N_{dof}$ & $\chi^2_r$ & Odds\\
\cmidrule(l){1-8}
A & $-1.0$ & $<16.2$ & 0.14 & 0.37 & 10 & 1.08 & $10^{-0.64}$\\
 & $-1.7$ & $<7.4$ & 0.14 & 0.38 & 10 & 1.16 & $10^{-1.05}$\\
 & $-2.5$ & $<2.2$ & 0.14 & 0.38 & 10 & 1.17 & $10^{-1.35}$\\
 & & & & & & & \\
B-0 & $-1.0$ & $5.3^{+2.5}_{-5.3}$ & 0.17 & 0.45 & 10 & 0.79 & $10^{-0.55}$\\
 & $-1.7$ & $<4.9$ & 0.15 & 0.53 & 10 & 1.07 & $10^{-1.21}$\\
 & $-2.5$ & $<0.9$ & 0.13 & 0.47 & 10 & 1.27 & $10^{-1.42}$\\
 & & & & & & & \\
B-1 & $-1.0$ & $<3.8$ & 0.14 & 0.40 & 10 & 1.06 & $10^{-0.85}$\\
 & $-1.7$ & $<1.5$ & 0.13 & 0.41 & 10 & 1.14 & $10^{-1.03}$\\
 & $-2.5$ & $<0.4$ & 0.14 & 0.38 & 10 & 1.17 & $10^{-1.22}$\\
 & & & & & & & \\
B-2 & $-1.0$ & $<2.4$ & 0.14 & 0.36 & 10 & 1.07 & $10^{-1.02}$\\
 & $-1.7$ & $<1.2$ & 0.14 & 0.38 & 10 & 1.13 & $10^{-1.00}$\\
 & $-2.5$ & $<0.4$ & 0.14 & 0.38 & 10 & 1.17 & $10^{-1.23}$\\
 & & & & & & & \\
C & $-1.0$ & $1.4^{+0.5}_{-0.7}$ & 0.25 & 0.40 & 10 & 0.55 & $10^{+0.28}$\\
 & $-1.7$ & $1.1\pm 0.4$ & 0.21 & 0.47 & 10 & 0.59 & $10^{+0.12}$\\
 & $-2.5$ & $1.1^{+0.4}_{-0.6}$ & 0.17 & 0.58 & 10 & 0.75 & $10^{-0.46}$\\
 & & & & & & & \\
Null\tmark[b] & $\dots$ & 0 & 0.15 & 0.26 & 11 & 0.81 & 1.0\\
\bottomrule
\end{tabular}
\begin{tablenotes}
\footnotesize
\item[a] Best fit model parameters for maximum likelihood (ML) number of pulsars.  If the ML number of pulsars is zero, then the 68\% confidence upper limit is used and reported with a ``$<$''.
\item[b] The ``Null'' case fixes the number of pulsars at zero.
\end{tablenotes}
\end{threeparttable}}
%%\end{center}

\section{Diffuse Gamma-Ray Emission and MSPs}
\label{diffuse_gamma}
MSPs are known gamma-ray sources \citep{abdo09}. As a result, we may set constraints on the MSP population in the GC by measuring the diffuse gamma-ray emission from the GC.  In a recent analysis of the first two years of data from the Fermi Gamma-Ray Space Telescope, \citet[herafter, HG]{hg11} observed an excess of gamma-ray flux toward the inner $1^\circ$ (150~pc) of the GC, with a significant excess within $0.25^\circ$ ($\approx40$~pc).  HG argued that the signal is consistent with the annihilation of 7--10 GeV dark matter particles with a cusped halo distribution around Sgr~A* and difficult to explain using known astrophysical sources.  \citet{aba11}, however, claimed that while the spectrum may be inconsistent with the \emph{average} pulsar spectrum, it is consistent with \emph{some} pulsar spectra and therefore could be explained by astrophysical sources.  HG provide a spectrum for the gamma-ray excess in the GC of the form
\begin{equation}
\frac{dN_\gamma}{dE} \propto E^{-\Gamma} \exp \left( -E / E_{cut}\right)
\end{equation}
where $\Gamma = 0.99^{+0.10}_{-0.09}$ and $E_{cut} = 1.92^{+0.21}_{-0.17}$~GeV.   These parameter values are consistent (to current uncertainties) with the spectra of 16 out of 46 pulsars in the first Fermi LAT catalog of gamma-ray pulsars \citep{abdoPSR10}.  We proceed assuming that the observed excess seen by HG is real \citep[however, see ][]{bmr11}, follows the spectrum fit by HG and is entirely caused by a collection of MSPs in the GC. 

The spectrum is normalized so that 
$E^2 dN_\gamma/dE \approx 10^{-7}~\mbox{GeV}~\mbox{cm}^{-2}~\mbox{s}^{-1}$ 
at $E=1.0$~GeV based on the HG plots, giving
\begin{eqnarray}
\frac{dN_\gamma}{dE} &=& 1.7 \times 10^{-7} \mbox{ GeV}^{-1} \mbox{cm}^{-2} \mbox{s}^{-1} \left(\frac{E}{1 \mbox{ GeV}}\right)^{-0.99} \nonumber \\
&  &\times\exp{\left( -\frac{E}{1.92 \mbox{ GeV}} \right)}.
\end{eqnarray}

Calculating the integrated energy flux from the above spectrum over 0.1--100~GeV as
\be
S_{\gamma} = \int_{0.1~\mbox{\scriptsize GeV}}^{100~\mbox{\scriptsize GeV}} E \frac{dN_\gamma}{dE} dE
\ee
and letting $L_{\gamma} = 4\pi d^2 f_{\Omega} S_{\gamma}$, we find that the gamma-ray luminosity of a source at the GC is $L_{\gamma} \approx 4 \times 10^{36} \mbox{ erg s}^{-1} f_{\Omega}$.  The correction factor, $f_{\Omega}$, is similar to the beaming fraction in radio pulsars and is generally taken to be unity in most modern models \citep{wrw09}.

Following calculations made to estimate the number of MSPs in globular clusters \citep{abdoGC10}, we can estimate the MSP population of the GC as 
\begin{equation}
\label{gamma_est}
\nmsp = \frac{L_{\gamma}}{\langle \dot{E} \rangle \langle \eta_{\gamma} \rangle} 
\end{equation}

where $\dot{E}$ is the total spindown luminosity of a MSP, $\eta_{\gamma} = L_{\gamma}/\dot{E}$ is the ``efficiency'' of converting spin-down power into gamma-rays, and angled brackets denote an average over the population.  
The values of $\langle \dot{E} \rangle$ and $\langle \eta_{\gamma} \rangle$ are taken from the local ($d<1$~kpc) MSP population to avoid selection effects.  
The mean spin-down luminosity is taken to be $\langle \dot{E} \rangle = 1.1 \times 10^{34} \mbox{ erg s}^{-1}$ from the 27 MSPs within 1~kpc of Earth listed in the ATNF catalog \citep{PSRCAT}.
Since 
$\langle \eta_{\gamma} \rangle \propto L_\gamma \propto d^2$, 
the uncertainties in $\langle  \eta_{\gamma}  \rangle$ are dominated by distance uncertainties and values range from $\langle  \eta_{\gamma}  \rangle = 10^{-3}-1$ in the first \emph{Fermi} gamma-ray pulsar catalog \citep{abdoPSR10}.  
To mitigate this distance problem, we take only those seven nearby pulsars in the catalog for which distances could be measured accurately with parallax.  
For these pulsars, we find $\langle \eta_{\gamma} \rangle = 0.08\pm 0.04$.
Using Equation~\ref{gamma_est}, we get an estimate of $\nmsp \approx 5000$.  
In order to contribute the observed excess, these pulsars would be located within $1^{\circ}$ ($150$ pc) of Sgr~A*, with the highest concentration within $0.25^{\circ}$ ($40$ pc).

Our estimate for the number of MSPs in the GC is essentially an upper bound for the population.  However, modeling the background component of the diffuse gamma-ray emission in the GC is still somewhat uncertain \citep{abdoSRC09} and, as a result of this uncertainty, our estimate can only be taken as an approximate upper limit.  Additionally, as this is a measure of the excess gamma-ray flux in the region (with the Galactic plane and a central point source coincident with \sgra\ subtracted), there exists the possiblity that a signicant number of MSPs are unaccounted for in this estimate.  Assuming the HG gamma-ray excess in the GC is real and does not suffer from systematic errors in background subtraction, we see that it is not inconsistent with a centrally concentrated population of ${\sim}10^3$ MSPs in the inner tens of parsecs from Sgr~A*.      

Finally, we note that similar estimates for the MSP population in the GC have been made in the past. \citet{wjc05} used a model proposed by \citet{zc97} to predict the emission of gamma-rays from MSPs as a function of global pulsar parameters like spin period and polar magnetic field.  They used Monte Carlo methods to simulate a population of pulsars and measured the total gamma-ray luminosity.  Comparing this luminosity to measured values from EGRET, \citet{wjc05} estimated a GC MSP population of $\nmsp\approx6000$ in the EGRET field ($r \approx 1.5^{\circ}$, 220~pc).

\section{Massive Stars in the Galactic Center}
\label{massive_stars}
The central parsec of the GC is one of the most active massive star formation regions in the Milky Way and is currently known to contain about 200 young massive stars \citep{geg10}.  As massive stars are the progenitors of neutron stars (NSs), we may use current stellar populations to estimate the number of pulsars in this region.

The inner parsec stellar population is divided into three fairly distinct regions.  The innermost region $\left(R\leq1^{\prime\prime}\right)$ is the so-called ``S-star Cluster'' of main sequence B-stars.  These stars can be fitted with a standard Salpeter initial mass function (IMF) for a single star formation event or for a continuous star-forming population with ages of a few Myr to 60 Myr \citep{bartko10}.  Outside this region $\left( 1^{\prime\prime}\leq R\leq 12^{\prime\prime}\right)$, the stars are largely arranged in at least one disk (possibly two) of mass $M\sim 10^4 M_{\odot}$ with a top-heavy IMF of $dN/dm \propto m^{-0.45}$ \citep{bartko10}.  The early-type stars in this region appear to have been formed in a starburst ${\sim}6$ Myr ago.  Outside the disk region $\left( R \geq 12^{\prime\prime}\right)$ the stellar population is again consistent with a Salpeter IMF.

Following similar calculations by \citet{lc08} and \citet{fl11}, the current stellar populations can be used to estimate the number of radio pulsars harbored in the central parsec.  The number of active CPs and MSPs beamed toward Earth are estimated to be
\be
\label{eq:cppops}
\ncp = f_{psr} f_{b} f_{\tau}  f_{v} N_{ns}
\ee   
and
\be
\label{eq:msppops}
\nmsp = f_{psr} f_{b} f_{\tau} f_{v} f_{r} N_{ns},
\ee 
respectively, where $N_{ns}$ is the number of NSs in the inner parsec, $f_{psr}$ is the fraction of NSs that form pulsars, $f_{b}=0.2$ is the beaming fraction, $f_{\tau}$ is the fraction of pulsars with ages less than the typical pulsar radio lifetime, $f_{v}$ is the fraction of pulsars with birth velocities small enough to be retained in the inner parsec and $f_{r}$ is the fraction of NSs that are recycled into MSPs.  The fraction of NSs that form pulsars is taken to be $f_{psr}\sim1$, although this factor is still fairly uncertain \citep{lc08}.  The terms $N_{ns}$, $f_{\tau}$, $f_{v}$ and $f_{r}$ are discussed below.

\subsection{Neutron Star Population in the Inner Parsec ($N_{ns}$)}  
The total number of NSs residing in the inner parsec may be estimated from current observations of the massive star population.  Velocity measurements of stars allow for the total dynamical mass enclosed within a few parsecs of {\sgra} to be determined.  Subtracting the mass of the MBH gives an extended mass of ${\sim}10^6~M_{\odot}$ in stars within a parsec of {\sgra} \citep{gtk96, sme09}.  With the total mass of stars established, the number of NSs can be calculated for a given initial mass function (IMF) and mass range of stars that end their lives as NSs.  

A typical NS progenitor mass range is $9M_{\odot}<M<25M_{\odot}$ \citep{hfw03}.  However, such ranges are theoretically determined for only isolated non-rotating stars and rely on mass-loss and stellar wind models that are poorly constrained by observations \citep{hfw03}.  If very massive stars have higher mass-loss rates than modeled, then NSs could form from stars with initial masses above $\gtrsim25M_{\odot}$.  That this may be the case is supported by the limited observational constraints.  For example, \citet{mcc06} detected an X-ray pulsar in the young Galactic cluster Westerlund~1 that requires a progenitor mass of $M>40M_{\odot}$ to have formed in the age of the cluster.

To calculate the number of NSs produced by the GC massive star population, we consider IMFs of the form $dN/dm\propto m^{-\alpha}$ over a range of masses from $0.1M_{\odot}$ to $100M_{\odot}$.  For the case of the standard Salpeter IMF ($\alpha=2.35$) observed in most of the inner parsec and a progenitor mass range of $9M_{\odot}<M<25M_{\odot}$, a total of $N_{ns}\approx4900$ NSs are produced.  If the top-heavy IMF ($\alpha=0.45$) observed in the young disks is taken to hold for the whole central parsec, a total of $N_{ns}\approx5700$ are produced.  If a wider range of progenitor masses is taken, say $9M_{\odot}<M<40M_{\odot}$ to allow the \citet{mcc06} observation, the total NS populations increase to $N_{ns}\approx5700$ and $N_{ns}\approx9500$ for the Salpeter and top-heavy IMFs, respectively.  
Even given a wide range in IMF and progenitor mass, the above results are consistent to an order of magnitude with a NS population of $N_{ns}\sim10^4$.  In terms of NSs formed per stellar mass, we find a value of $\beta_{ns}\sim10^{-2}M^{-1}_{\odot}$ for the ${\sim}10^6M_{\odot}$ worth of stars within a parsec of \sgra.

\subsection{Fraction of Still-Active Pulsars ($f_{\tau}$)}
\label{ss:f_r}

The typical lifetimes over which pulsars maintain active radio emission are $\tau\sim10^7$~yr for CPs and $\tau\sim10^9-10^{10}$~yr for MSPs.  As a result, the fraction of pulsars formed recently enough to still be active must be considered in estimates of the observable population.  This factor will depend on the star formation history (SFH) of the region.  In the case of continuous star formation, the active fraction of pulsars can be estimated to be $f_{\tau}\sim\tau/t_{sf}$, where $t_{sf}$ is the amount of time elapsed since the star formation began.

\subsection{Fraction of Pulsars Retained ($f_{v}$)}
\label{ss:fv}
From the high observed velocities (${\sim}10^2-10^3~\mbox{km~s}^{-1}$) of some pulsars, it has been inferred that NSs are given a large ``kick'' velocity at birth as a result of binary disassociation or an asymmetric supernova explosion (or both).  A pulsar created in the inner parsec will be retained in the inner parsec only if its birth velocity does not exceed the local escape velocity of its orbit around Sgr~A*.  Assuming only the influence of a MBH of mass $M\approx 4 \times 10^6~\mbox{M}_{\odot}$, the escape velocity at a distance $r$ from Sgr~A* is given by

\begin{equation}
\label{escape_vel}
v_e(r) = 185~\mbox{km s}^{-1}~\left(\frac{M}{4\times10^6~\mbox{M}_{\odot}}\right)^{1/2}\left(\frac{r}{1~\mbox{pc}}\right)^{-1/2}.
\end{equation}

Assuming a Maxwellian distribution of birth velocities with a mean of $\langle v_{birth} \rangle = 380~\mbox{km s}^{-1}$ \citep{fgk06}, we find retention fractions of $f_v \approx 0.1$ and $f_v \approx 0.25$ for radial distances of $r=1~\mbox{pc}$ and $r=0.5~\mbox{pc}$, respectively.  However, the actual shape of the pulsar birth velocity distribution is not well constrained and one may worry that the Maxwellian distribution is arbitrary.  In an analysis of pulsar velocities, \citet{fgk06} consider six different pulsar velocity distributions.  Repeating our calculation for each distribution, we find retention fractions in the ranges of $f_v \approx 0.05 - 0.4$ and $f_v \approx 0.1 - 0.5$ for distances of $r = 1~\mbox{pc}$ and $r=0.5~\mbox{pc}$, respectively.

If no other effects are important, we would expect a retention fraction of $f_v\gtrsim0.1$.  However, a similar analysis to the one performed above would underestimate the retained NS populations in globular clusters by orders of magnitude.  Observations of pulsars and low-mass X-ray binaries (LMXBs) suggest that up to ${\sim}10\%$ of all NSs formed in some globular clusters may be retained \citep{prp02}.  However, current models for isolated pulsars predict that $\lesssim1\%$ of NSs will have birth velocities below the $\lesssim50~\mbox{km s}^{-1}$ globular cluster escape velocities.  This ``retention problem'' is currently an unsolved problem, but likely has to do with the high stellar densities and binary fractions found in the cores of globular clusters \citep[see][and references within]{prp02}.  Since the GC has even higher stellar densities than cores of globular clusters, we expect a similar heightening of the retention fraction and thus adopt as a nominal value $f_v \sim 1$.  

\subsection{Fraction of Pulsars Recycled to MSPs ($f_{r}$)}
\label{ssec:f_r}
MSPs are thought to be formed when a NS in a binary gains angular momentum through accretion of matter in a process known as ``recycling'' \citep{acr82,bv91}.  Thus, the fraction of NSs recycled into MSPs must be determined to estimate the MSP population. Within 3~kpc of the Sun, the birthrate of pulsars with 400~MHz pseudo-luminosities above $1~\mbox{mJy}~\mbox{kpc}^2$ is observed to be ${\sim}10^{-3}~\mbox{yr}^{-1}$ and ${\gtrsim}10^{-6}~\mbox{yr}^{-1}$ for CPs and MSPs, respectively \citep{parkesII1997}.  Using these birthrates, we may infer that the recycling fraction is at least $f_r\gtrsim10^{-3}$ in the Galactic disk.  The increased stellar density and stellar encounter rate in the GC will very likely increase this fraction.  LMXBs, the assumed progenitors of MSPs, have been found to be ${\sim}100$ times more abundant in globular clusters than the general Galactic field \citep{clark75, katz75}.  As the central parsec of the GC has a higher stellar density than globular clusters, we would expect at least a similar overabundance of LMXBs and their resultant MSPs as is seen in globular clusters.  As a result, we adopt a recycling fraction of $f_r\sim0.1$.

\subsection{Pulsar Estimates for Various Star Formation Histories}

As the star formation history (SFH) of the central parsec of the GC is still somewhat uncertain, we will consider two general SFHs suggested by current observations.  In the first case, we take the massive-star disk(s) to have formed in a well-defined starburst ${\sim}6$~Myr ago \citep{pgm06,bartko10} and assume that the rest of the central parsec has experienced continuous star formation over the age of the Galaxy.  In the second case, we consider SFHs based on spectro-photometry of cool giant stars which indicate that most of the stars in the inner parsec were formed $\gtrsim5$~Gyr ago but also show an increased star formation rate in the last ${\sim}100$~Myr \citep{brs03, pfz11}. 

\subsubsection{Continuous Star Formation + Disk Starburst}

In the first SFH, pulsars can come from both the general population of stars in the central parsec and the young disk population.  For the general population, we take the total mass of stars to be $M\sim10^6{M}_\odot$ and assume continuous star formation over the last ${\sim}10^{10}$~yr.  From the parameters discussed above, the total number of NSs is found to be 
$N_{ns}\sim10^4\left(\beta_{ns}/10^{-2}\mbox{M}^{-1}_\odot\right)$.
Taking a CP active radio lifetime of $\tau\sim10^7$~yr, continuous star formation over ${\sim}10^{10}$~yr will give the fraction of CPs still active to be $f_{\tau}\sim10^{-3}$.  From Equation~\ref{eq:cppops}, the CP contribution from the continuous star forming region of the inner parsec is
\be
N^{\mbox{\scriptsize {gen}}}_{\mbox{\scriptsize {CP}}}\sim2\left(\frac{f_v}{1.0}\right)\left(\frac{\beta_{ns}}{0.01~\mbox{M}^{-1}_\odot}\right)\left(\frac{M}{10^6~\mbox{M}_\odot}\right).
\ee
For MSPs with a radio lifetime of ${\sim}10^{10}$~yr, the fraction still active is $f_{\tau}\sim1$.  Adopting the recycling fraction of $f_r\sim0.1$ discussed above, the contribution of MSPs from the general population of the inner parsec is given by Equation~\ref{eq:msppops} as
\be
N^{\mbox{\scriptsize {gen}}}_{\mbox{\scriptsize {MSP}}}{\sim}200\left(\frac{f_v}{1.0}\right)\left(\frac{f_{r}}{0.1}\right)\left(\frac{\beta_{ns}}{0.01~\mbox{M}^{-1}_\odot}\right)\left(\frac{M}{10^6~\mbox{M}_\odot}\right).
\ee

Additionally, the contribution of pulsars from the massive-star disk(s) must be considered.  The disk is assumed to have a population of stars formed in a starburst event ${\sim}6\times10^6$~yr ago with a total stellar mass of ${\sim}10^4{M}_{\odot}$  \citep{pgm06,bartko10}.  Since the age of the disk is comparable to the active radio lifetime of a CP, the fraction of CPs still active is taken to be $f_{\tau}\sim1$.  From Equation~\ref{eq:cppops}, the CP contribution from the disk is found to be
\be
N^{\mbox{\scriptsize {disk}}}_{\mbox{\scriptsize {CP}}}\sim20\left(\frac{f_v}{1.0}\right)\left(\frac{\beta_{ns}}{0.01~\mbox{M}^{-1}_\odot}\right)\left(\frac{M_{disk}}{10^4~\mbox{M}_\odot}\right).
\ee
The disk population is not expected to produce any currently observable MSPs as the short timescale of ${\sim}6\times10^6$~yr provides insufficient time to create and evolve a NS population into MSPs.  As a result, this first SFH produces roughly $\ncp\sim20$ CPs and $\nmsp\sim200$ MSPs.    
   
\subsubsection{SFH from Observations of Cool Giant Stars}
 The SFH has also been estimated by comparing simulated populations with the observed cool giant stars in the central parsec.  Such simulations allow the average star formation rate to be calculated as a function of look-back time for a few coarse time bins.  In two separate analyses, both \citet{brs03} and  \citet{pfz11} found that ${\gtrsim}80\%$ of the stellar mass in the central parsec was formed ${\gtrsim}5$~Gyr ago and that there has been increased star formation in the last ${\sim}100$~Myr.  In their best fit models, \citet{brs03} found an average star formation rate of ${\sim}3\times10^{-3}M_{\odot}\mbox{yr}^{-1}$ within 2~pc of {\sgra} from 10 to 100~Myr ago and \citet{pfz11} found an average star formation rate of ${\sim}10^{-3}M_{\odot}\mbox{yr}^{-1}$ within 1~pc of {\sgra} from 50 to 200~Myr ago.  Both cases are consistent with ${\sim}10^5M_{\odot}$ worth of stars being formed in the inner parsec of the GC in the last ${\sim}100$~Myr.  

If there has been continuous star formation in the last ${\sim}100$~Myr, then the fraction of still active CPs would be $f_{\tau}\sim0.1$.  Taking all other parameters as before, the continuous formation of  ${\sim}10^5M_{\odot}$ worth of stars over the last ${\sim}10^8$~yr would produce
\be
N^{\mbox{\scriptsize {con}}}_{\mbox{\scriptsize {CP}}}\sim20\left(\frac{f_v}{1.0}\right)\left(\frac{\beta_{ns}}{0.01~\mbox{M}^{-1}_\odot}\right)\left(\frac{M\left(t<10^8~\mbox{yr}\right)}{10^5~\mbox{M}_\odot}\right).
\ee
If the recent star formation all took place in the last ${\sim}10^7$~yr, then the fraction of CPs still active would be $f_{\tau}\sim1$.  In this case, the number of active CPs would be
\be
N^{\mbox{\scriptsize {burst}}}_{\mbox{\scriptsize {CP}}}{\sim}200\left(\frac{f_v}{1.0}\right)\left(\frac{\beta_{ns}}{0.01~\mbox{M}^{-1}_\odot}\right)\left(\frac{M\left(t<10^8~\mbox{yr}\right)}{10^5~\mbox{M}_\odot}\right).
\ee
In either of the above cases, the majority $\left({\gtrsim}90\%\right)$ of the star formation took place at look-back times ${\gtrsim}10^8$~yr ago.  As a result, the number of MSPs produced will be approximately the same as the first SFH considered, namely $\nmsp\sim200$.

\subsubsection{Upper Limits to the Pulsar Population}
Using the above estimates of CP and MSP populations for a range of observationally supported SFHs, upper limits may be set on the total allowable number of pulsars in the inner parsec.  For CPs, the most favorable formation scenarios produce $\ncp\sim200$.  For MSPs, a variety of SFHs consistently produce $\nmsp\sim200\left(f_r/0.1\right)$.  Since the recycling fraction is unknown for the extreme conditions of the inner GC, an upper limit of $\nmsp\sim2000$ may be set by adopting $f_r\sim1$.  Thus, observations of current stellar populations place an upper limit of $\mbox{a few}\times10^3$ on the number of active radio pulsars beamed toward Earth in the inner parsec of the GC.

\section{Pulsar Wind Nebulae in Inner 20 pc}
\label{PWN_sec}
Using a total of 1~Ms of \emph{Chandra} ACIS--I observations of the inner parsecs of the GC, \citet{mbb08} compiled a catalog of 34 diffuse X-ray emitting features.  Based on the X-ray luminosities and sizes of the sources in their catalog, \citet{mbb08} expect ${\sim}20$ pulsar wind nebulae (PWNe) to be present within 20~pc of \sgra.  Since PWNe are powered by pulsars, we may use this inferred population of PWNe to estimate the pulsar population in the inner 20~pc of the GC.

Pulsars can lose their rotational kinetic energy by the release of relativistic winds of charged particles.  The winds exert a pressure upon and deposit energy into the surrounding interstellar medium, producing luminous PWNe that radiate across the electromagnetic spectrum \citep[see, e.g.,][]{gs06}.  As a result, the luminosity of the PWN will be directly related to the spin-down luminosity of the pulsar given by
\begin{equation}
\label{edot}
\dot{E} = -dE_{rot}/dt = 4\pi^2I\dot{P}/P^3,
\end{equation}  
where \emph{I} and \emph{P} are the moment of inertia and period, respectively, of the pulsar.  

Since the spin-down luminosity of a pulsar decreases with increasing age, one would expect the most luminous PWNe to contain young pulsars \citep[however, older pulsars may be ``recycled'' sufficiently to power PWNe as described by][]{ctw06}.  Of the 30 confirmed pulsars associated with PWNe, 25 have characteristic ages ($\tau_c = P/2\dot{P}$) of $\tau_c \lesssim 10^6~\mbox{yr}$ and 23 have characteristic ages $\tau_c \lesssim 10^5~\mbox{yr}$ \citep{roberts04}.  Thus, we adopt a lifetime for a typical PWN of $t_{pwn} \sim 10^5~\mbox{yr}$.

Given the observed number of PWNe in the GC ($N_{obs} \sim20$) and a typical lifetime of $t_{pwn} \sim 10^5~\mbox{yr}$, we find a mean rate of formation of PWNe over the last ${\sim}10^5~\mbox{yr}$ to be
\begin{equation}
\label{pwn_rate}
\beta_{pwn} \sim 2\times10^{-4}~\mbox{yr}^{-1}\left(\frac{N_{pwn}}{20}\right)\left(\frac{t_{pwn}}{10^5~\mbox{yr}}\right)^{-1}.
\end{equation}

Assuming that the PWN formation rate has remained constant over the last ${\sim}10^7~\mbox{yr}$, we may estimate the number of CPs in the inner 20~pc to be
\begin{equation}
\label{PWN_cp}
\ncp \sim \beta_{pwn} \tau_{psr} f_b f_v f_{pwn}^{-1},
\end{equation}
where $\tau_{psr}\sim10^7~\mbox{yr}$ is the typical radio lifetime of a CP, $f_b=0.2$ is the beaming fraction, $f_v$ is the fraction of pulsars with birth velocities low enough to be retained in the inner 20~pc, and $f_{pwn}\sim1$ is the fraction of pulsars that form PWNe.  

Taking the mass of the central 20~pc to be $M=3\times10^7~M_{\odot}$ \citep{lhw92}, we find that pulsars must have velocities $v_{birth} < 115~\mbox{km~s}^{-1}$ to remain gravitationally bound to the inner 20~pc.  From the birth velocity distributions considered by \citet{fgk06}, we find $f_v\approx0.05-0.30$.  However, the distance traveled by a pulsar is given by
\begin{equation}
\label{dist_psr}
d \approx 10~\mbox{pc}\left(\frac{v}{100~\mbox{km s}^{-1}}\right)\left(\frac{t}{10^5~\mbox{yr}}\right).
\end{equation}
Thus, pulsars with velocities high enough to become gravitationally unbound will also have velocities high enough to escape the inner 20~pc on timescales comparable to the PWN lifetime.  As a result, the PWNe observed within the inner 20~pc are very likely to remain there and we can take $f_v\sim1$.

From Equation~\ref{PWN_cp}, we see that if the PWN birth rate has been constant over the last ${\sim}10^7~\mbox{yr}$, we would expect
\begin{equation}
\label{npsr_pwn}
\ncp \sim 400 \left(\frac{N_{pwn}}{20}\right)\left(\frac{t_{pwn}}{10^5~\mbox{yr}}\right)^{-1}\left(\frac{f_{pwn}}{1.0}\right)^{-1}
\end{equation}
CPs within 20~pc of \sgra.

Finally, we note that of the ${\sim}20$ PWN candidates identified by \citet{mbb08}, 4 fall within a projected radial distance of 1~pc from \sgra.  Assuming the number of pulsars scales accordingly, then Equation~\ref{npsr_pwn} predicts $N_{CP}\sim80$ CPs within the inner parsec of the GC.

\section{Supernova Rate in the Galactic Center}
\label{supernova}
Neutron stars are formed as the end products of core-collapse supernovae (CCSN). An estimate of the rate of CCSN in the GC would therefore offer a constraint on the pulsar population.  The CCSN rate is estimated below for both $r<150~\mbox{pc}$ and $r<20~\mbox{pc}$.

\subsection{CCSN Rate Within $r<150~\mbox{pc}$ of \sgra}

By measuring the total mass of ${}^{26}\mbox{Al}$ in the Galaxy, \citet{diehl06} estimate the Galactic CCSN rate to be ${\beta}_{CCSN} = 1.9 \pm 1.1 {\mbox{ century}}^{-1}$. 
One may, in principle, scale this estimate to smaller regions of the Galaxy using massive star populations. Taking the inner 500 pc to contain 10\% of the Galaxy's massive star formation \citep{figer08} we can estimate that the inner ${\sim}150$ pc contains ${\sim}2\%$ of the massive star formation and therefore should have a CCSN rate of 
${\beta}_{CCSN} \approx 0.04 {\mbox{ century}}^{-1}$.  \citet{crocker11} estimate a similar rate and show that it is consistent with SN rate estimates from infrared observations, stellar composition, X-ray emission, gas turbulence, and high-velocity compact clouds \citep[see][and references within]{crocker11}. 
We may now estimate the CP population in the GC to be 
\be
\ncp = f_{psr}f_{b}f_{v}{\tau}_{psr}{\beta}_{CCSN}
\ee
where 
${\beta}_{CCSN} \approx 4 \times 10^{-4} {\mbox{yr}}^{-1}$ is the CCSN rate, 
${\tau}_{psr} \sim 10^7 \mbox{ yr}$ is the mean canonical pulsar lifetime, 
$f_b=0.2$ is the fraction of pulsars beamed toward Earth, 
$f_{v}$ is the fraction of pulsars with birth velocities small enough to be retained by the GC and $f_{psr}\sim1$ is the fraction of CCSN that result in active pulsars.  Using the distributions from \citet{fgk06}, the fraction of pulsars with birth velocities smaller than the escape velocity $v_e\approx200~\mbox{km s}^{-1}$ at 150~pc ranges from $f_v \approx 0.1-0.4$.  
These values give an estimate of $\ncp \sim100$.  Likewise, accounting for the longer ages for MSPs ($\tau_{psr}\sim10^{10}$~yr), we can estimate the MSP population to be $\nmsp\sim10^5f_r$, where $f_r$ is the fraction of NSs that get recycled to MSPs (see Section~\ref{ssec:f_r}).

\subsection{CCSN Rate in Inner 20~pc from X-ray Observations}

Studies of diffuse X-ray emission can also provide insight into the SN rate in the GC. Using over 600~ks of \emph{Chandra} ACIS--I observations, \citet{mbb04} found that the diffuse X-ray emissions in the GC could be explained by a two-temperature plasma composed of a ``soft'' component ($kT\approx0.8~\mbox{keV}$) and a ``hard'' component ($kT\approx8~\mbox{keV}$).  Assuming the soft component of the plasma is primarily heated by SNe, an estimate for the SN rate can be made by observing the loss of energy from the inner 20~pc.

Let us first consider the case in which the soft component of the plasma just cools radiatively.  The X-ray luminosity of the soft component of the plasma in the inner 20~pc is $L_X\approx3\times10^{36}~\mbox{erg s}^{-1}$ \citep{mbb04}.  If each SN transfers ${\sim}1\%$ of its total kinetic energy of ${\sim}10^{51}~\mbox{erg}$ to the plasma, then a SN rate of $\beta_{SN}\approx10^{-5}~\mbox{yr}^{-1}$ is required to maintain the currently observed temperature.  Taking this rate to be constant and $f_v\sim0.1$ (see Section~\ref{PWN_sec}) gives an estimate for the CP population of $\ncp\sim2\left(f_v/0.1\right)$.

If the plasma is unconfined, it can also cool through adiabatic expansion.  Rough estimates put this cooling rate at $L_{ad}\approx9\times10^{38}~\mbox{erg s}^{-1}$ \citep{mbb04}, which would require a SN rate of $\beta_{SN}\approx3\times10^{-3}~\mbox{yr}^{-1}$.  Again assuming this rate is constant and $f_v\sim0.1$ as above, the estimated CP population is $\ncp\sim600\left(f_v/0.1\right)$.

The above estimates assume that the SN rate is constant over the radio lifetime of a CP (${\sim}10^7$~yr).  However, the SN rate estimates are only required to hold over a characteristic cooling time of $t_{c}\sim E/L$, where $E\sim5\times10^{50}~\mbox{erg s}^{-1}$ is the total thermal energy stored in the plasma and $L$ is the appropriate cooling luminosity \citep{mbb04}.  The cooling timescales for radiative cooling and adiabatic expansion are $5\times10^6$~yr and $2\times10^4$~yr, respectively.  

Keeping this caveat in mind, the assumption of a constant SN rate does allow useful bounds to be put on the pulsar populations in the GC.  Taking a constant SN rate from radiative cooling to be a lower bound and a constant SN rate from adiabatic expansion to be an upper bound, we find the CP population in the inner 20~pc to be $1\lesssim \ncp\lesssim10^3$.  Likewise, for MSPs with $\tau\sim10^{10}$~yr, we estimate a population of $10^3f_r\lesssim \nmsp\lesssim10^6f_r$, where $f_r$ is the fraction of NSs recycled into MSPs (see Section~\ref{ssec:f_r}).

\section{Discussion}
\label{discussion}
\subsection{Summary of Estimates}
We have used observations over a wide range of wavelengths to make order of magnitude estimates of the number of pulsars allowed within $r\leq1$~pc and $r\leq150$~pc of Sgr~A*.  The estimates are summarized in Table~\ref{summary1}.  

\begin{table*}[h!t]
\begin{center}
\begin{threeparttable}
\caption{Summary of Pulsar Population Estimates by Method} 
\label{summary1}
\begin{tabular}{lcccccc}
\toprule
\toprule
& \multicolumn{3}{c}{$r\leq1~\mbox{pc}$} & \multicolumn{3}{c}{$r\leq150~\mbox{pc}$}\\
\cmidrule(l){2-4}  \cmidrule(l){5-7}
Method & CP & MSP & Total & CP & MSP & Total\\
\midrule
Pulsar Surveys & ${\lesssim}10^3$ & $\dots$ & $\dots$  & $\dots$ & $\dots$ & ${\gtrsim}10^2$\\
Radio Point Sources  & $\dots$ & $\dots$ & $\dots$ & $\dots$ & $\dots$ & ${\lesssim}10^4$\\
Radio Spectrum A& $\dots$ & $\dots$ & $< 7.4\times10^3$ & $\dots$ & $\dots$ & $\dots$\\
Radio Spectrum B-1& $\dots$ & $\dots$ &$<1.5\times10^3$ & $\dots$ & $\dots$ & $\dots$\\
Radio Spectrum C& $\dots$ & $\dots$ &$(1.1\pm0.4)\times10^3$ & $\dots$ & $\dots$ & $\dots$ \\
Diffuse Gamma-ray  & $\dots$ & $\dots$ & $\dots$ & $\dots$ &${\sim}5\times10^3 $& $\dots$ \\
Massive Stars\tmark[a] &${\lesssim}10^2$ & ${\lesssim}10^3$ & ${\lesssim}10^3$ & $\dots$ & $\dots$ & $\dots$\\
PWNe & $10^2$ & $\dots$ & $\dots$ & $\dots$ & $\dots$ & $\dots$ \\
Supernovae\tmark[a,b]&${\lesssim}10^3$ & ${\lesssim}10^4(f_{r}/10^{-2})$ & ${\lesssim}10^4$& $10^2$ & $10^3\left(f_{r}/10^{-2}\right)$ & $\dots$\\
\bottomrule
\end{tabular}
\begin{tablenotes}
\footnotesize
\item[a] Typical values for the ``recycling fraction'' are $f_r\gtrsim10^{-3}$ in the Galactic field and potentially as high as $f_r\gtrsim0.1$ in the central parsec (see Section~\ref{ssec:f_r})
\item[b] Supernovae estimate given at $r\leq20$~pc and $r\leq150$~pc
\end{tablenotes}
\end{threeparttable}
\end{center}
\end{table*}

\subsubsection{Pulsar Population Within $r\leq150$~pc}
Limits on the pulsar population within 150~pc of \sgra\ may be set using the five known pulsars in the inner $15\arcmin$, the catalog of compact radio sources by \citet{lc08}, measurements of an excess gamma-ray flux by \citet{hg11} and estimates of the Galactic core-collapse supernova rate.  

The five known pulsars in the inner $15\arcmin$ provide the strongest current evidence for an intrinsic pulsar population in the GC region.  A Monte Carlo population analysis by \citet{dcl09} showed that the pulsar detections in the survey by \citet{dcl09} indicate a population of at least $N\gtrsim100$ pulsars within 100~pc of \sgra.  

The catalog of compact radio sources compiled by \citet{lc08} allows for upper limits to be placed on the pulsar population in the inner 150~pc.  The VLA survey produced a total of 170 compact steep-spectrum sources.  Although pulsars cannot be unequivocally classified in an imaging survey, \citet{lc08} estimate that $N\sim10$ of the sources were likely pulsars.  Using this estimate and the survey sensitivities, a conservative upper bound of $N\lesssim10^4$ may be set on the pulsar population in this region.  Even if it turns out that there are no pulsars in the catalog, an upper limit of $N\lesssim10^3$ may be set to 99\% confidence level.  We note that although these upper limits should hold for the entire survey region, the increasing background temperature and sidelobes caused by extended emission toward the inner GC mean that the inner field of the survey (half-power radius of 15\arcmin) likely experienced decreased sensitivity and could potentially hide a significant pulsar population in the immediate vicinity of \sgra.

Using the first two years of \emph{Fermi} data, \citet{hg11} claim to have detected an excess diffuse gamma-ray flux in the inner 150~pc of the Galaxy, which they attribute to annihilating dark matter particles.  If we assume the gamma-rays come instead from a collection of MSPs, an upper limit to the number of MSPs in the GC may be set.  Following similar calculations for globular clusters by \citet{abdoGC10}, we find that the excess is consistent with a population of ${\sim}5\times10^3$ MSPs.  This estimate is nominally an upper limit, but considering the systematic uncertainties in current GC gamma-ray background models and difficulties subtracting the point source associated with \sgra, we adopt this value as only a lower limit on the upper bound of MSPs in the inner 150~pc.

As radio pulsars are formed in core-collapse SNe, the SN rate also provides a constraint on the pulsar population in the GC.  By scaling down the Galactic SN rate based on massive star populations in a manner similar to that of \citet{crocker11}, we estimate a SN of $\beta_{CCSN}\approx0.04~\mbox{century}^{-1}$.  Such a SN rate 
indicates a CP population of $\ncp \sim100$ and an MSP population of $\nmsp \sim10^5f_r$, where $f_r$ is the recycling fraction discussed in Section~\ref{ss:f_r}.  Since the recycling fraction of field pulsars is $f_r\gtrsim10^{-3}$ and potentially as high as $f_r\sim0.1$ in globular clusters, we adopt a nominal value of $f_r\sim10^{-2}$, so $\nmsp \sim10^3 (f_r/10^{-2})$.

Finally, we note that all of the upper limits considered for pulsar populations on the large scale of 150~pc may not include contributions from the few inner parsecs as a result of decreased sensitivity or failure to incorporate recent starburst activity.  As a result, we consider these limits mainly applicable in the region $1~\mbox{pc}\lesssim r \lesssim 150~\mbox{pc}$.  The above estimates are consistent with a total pulsar population (that is, both CPs and MSPs) of $10^2\lesssim N_{psr} \lesssim 10^4$, a CP population of $\ncp \lesssim 10^2$ and an MSP population of $\nmsp \lesssim 10^4$.  
 
\subsubsection{Pulsar Population Within $r\leq1$~pc}

Limits on the pulsar population in the inner parsec of the GC have been set using the non-detections of high frequency directed pulsar searches, the spectrum of \sgra\ on arcsecond scales, the population and star-formation history of the massive star progenitors of NSs, the observations of ${\sim}20$ PWN candidates in the inner 20~pc, and the limits on the SN rate based on X-ray observations in the inner 20~pc.  

Directed pulsar searches of the inner parsec of the Galaxy have been conducted with the GBT at frequencies of 5, 9 and 15 GHz \citep{dcl09, mkfr10}.  Since none of these searches made any detections, an upper limit may be set on the total pulsar population that is still consistent with a null result.  Using the survey parameters provided in the \citet{dcl09} and \citet{mkfr10} surveys, we find that up to $N_{psr}\lesssim10^3$ pulsars (both CPs and MSPs) may be present in the inner parsec.

Since the compact radio source \sgra\ is broadened by interstellar scattering (${\approx}1\arcsec$ at 1~GHz), the observed flux density may actually be a combination of the emission near the MBH and a diffuse component from a population of unresolved pulsars.  By requiring that this two-component system reproduce the observed spectrum of \sgra, constraints may be placed on the pulsar population on arcsecond scales.  We consider a variety of spatial distributions and find that a total population of ${\sim}10^3$ pulsars is consistent with flux density measurements, regardless of spatial distribution.  The existence of such a large pulsar population would distort the low-frequency measurements of the intrinsic spectrum of \sgra\ and the free-free absorption along the line of sight of \sgra.

Upper limits to the total pulsar population can also be set by studying the populations of massive stars that end their lives as NSs.  Infrared observations of the present day population of massive stars allows for estimates of the star formation history.  We consider the case of two general SFHs and find that the CP population can get as high as $\ncp\lesssim200$ only under the most favorable conditions.  More typical estimates for the CP population are $\ncp \sim 20$, with most of these being formed in the young disk of massive stars located ${\approx}0.5$~pc from \sgra.  The MSPs are less sensitive to the exact SFH and produce populations of $\nmsp\sim200$ with an upper limit of $\nmsp\lesssim2000$ for a range of reasonable SFHs.

Upper limits on the number of CPs in the inner parsecs of the GC may be set using the detection of ${\sim}20$ PWNe within 20~pc of \sgra.  Using the ${\sim}20$ PWN candidates compiled by \citet{mbb08} in a catalog of diffuse X-ray sources and assuming PWNe are produced at a constant rate, we find that as many as 400 CPs may reside in the inner 20~pc.  If the CP distribution follows that of the PWN candidates, then as many as 80 CPs could reside in the inner parsec.  

Finally, we consider measurements of the soft ($kT\approx0.8$~keV) component of the diffuse X-ray plasma in the inner 20~pc.  Assuming that the plasma was heated by the transfer of kinetic energy from SNe, we may set an upper limit on the number of pulsars in this region \citep{mbb04}.  By considering two different cooling regimes, we find that the most extreme cooling scenario will produce a population of up to ${\sim}10^3$ CPs and ${\sim}10^6f_r$ MSPs.  Thus, we can set upper bounds of $\ncp\lesssim10^3$ and $\nmsp \lesssim10^4(f_r/10^{-2})$ for CPs and MSPs in the inner 20~pc.  

Overall, we find that the above estimates are consistent with a CP population of $\ncp \lesssim100$ and an MSP population of $\nmsp \lesssim 10^3$ in the inner parsec of the GC.

\subsection{Conclusions}
Current observations of the GC are consistent with a population of up to ${\sim}10^3$ active pulsars beamed toward Earth within the central parsec around Sgr~A*.  
This total population may consist of up to $\ncp \lesssim 100$ CPs and as many as $\nmsp \lesssim 10^3$ MSPs.  
Such a population could distort the low-frequency measurements of both the spectrum of \sgra\ and free-free absorption along the line of sight of \sgra.  
However, even with a potentially sizeable collection of pulsars, the difficult observing conditions of the inner regions of the GC will make individual detections a challenge.  
The strong interstellar scattering and large pulse broadening times mean that typical pulsar periodicity searches at radio wavelengths will be sensitive to pulsars only at high observing frequencies and large bandwidths \citep{cl97}.  
Even if most of the pulsars lie below the detection threshold, ``giant'' pulses intrinsic to the pulsar or as the result of an enhancement through multipath scattering may make a pulsar visible a search for single pulses.  Since there may be many more MSPs than CPs, one may search for pulsars using interferometer imaging surveys of compact radio objects similar to that of \citet{lc08}.  
Finally, search methods should be considered at wavelengths less affected by the scattering effects of the ISM.  
Since many MSPs produce gamma-ray emission, a directed search of the GC region with the \emph{Fermi} LAT could potentially detect a pulsar in a blind periodicity search. 

Despite the difficulties in finding pulsars in the GC, the detection of a pulsar orbiting \sgra\ with an orbital period of $P_{orb}\lesssim100$~yr would provide an unparalleled test of gravity in the strong-field regime and could potentially allow the measurement of the spin and quadrupole moment of the MBH \citep{pl04, lw97, wk99, lwk12}.  Additionally, the detection of even one pulsar in the inner few parsecs would provide an excellent probe of the magneto-ionic material, the gravitational potential, and the star formation history in the vicinity of Sgr~A*.  In light of the significant scientific rewards and a potentially sizeable target population, we strongly recommend continued pulsar searches in the GC across a wide range of wavelengths.   

\acknowledgments
We thank an anonymous referee for helpful comments that improved the clarity of this paper.  This research was supported at Cornell University by NSF grants AST-1008213 and AST-1109411.  Part of this research was conducted at the Jet Propulsion Laboratory, California Institute of Technology, under contract with the National Aeronautics \& Space Administration.  We have also made use of NASA's Astrophysics Data System.

\bibliography{rsw}

\begin{thebibliography}{76}
\expandafter\ifx\csname natexlab\endcsname\relax\def\natexlab#1{#1}\fi

\bibitem[{{Abazajian}(2011)}]{aba11}
{Abazajian}, K.~N. 2011, {Journal of Cosmology and Astroparticle Physics}, 3,
  10

\bibitem[{{Abdo} {et~al.}(2009{\natexlab{a}}){Abdo}, {Ackermann}, {Ajello},
  {et~al.}}]{abdo09}
{Abdo}, A.~A., {Ackermann}, M., {Ajello}, M., {et~al.} 2009{\natexlab{a}},
  Science, 325, 848

\bibitem[{{Abdo} {et~al.}(2009{\natexlab{b}}){Abdo}, {Ackermann}, {Ajello},
  {et~al.}}]{abdoSRC09}
---. 2009{\natexlab{b}}, \apjs, 183, 46

\bibitem[{{Abdo} {et~al.}(2010{\natexlab{a}}){Abdo}, {Ackermann}, {Ajello},
  {et~al.}}]{abdoGC10}
---. 2010{\natexlab{a}}, \aap, 524, A75

\bibitem[{{Abdo} {et~al.}(2010{\natexlab{b}}){Abdo}, {Ackermann}, {Ajello},
  {et~al.}}]{abdoPSR10}
---. 2010{\natexlab{b}}, \apjs, 187, 460

\bibitem[{{Alpar} {et~al.}(1982){Alpar}, {Cheng}, {Ruderman}, \&
  {Shaham}}]{acr82}
{Alpar}, M.~A., {Cheng}, A.~F., {Ruderman}, M.~A., \& {Shaham}, J. 1982, \nat,
  300, 728

\bibitem[{{An} {et~al.}(2005){An}, {Goss}, {Zhao}, {et~al.}}]{an05}
{An}, T., {Goss}, W.~M., {Zhao}, J.-H., {et~al.} 2005, \apjl, 634, L49

\bibitem[{{Bartko} {et~al.}(2010){Bartko}, {Martins}, {Trippe},
  {et~al.}}]{bartko10}
{Bartko}, H., {Martins}, F., {Trippe}, S., {et~al.} 2010, \apj, 708, 834

\bibitem[{{Bates} {et~al.}(2011){Bates}, {Johnston}, {Lorimer}, {Kramer},
  {et~al.}}]{bjl11}
{Bates}, S.~D., {Johnston}, S., {Lorimer}, D.~R., {Kramer}, M., {et~al.} 2011,
  \mnras, 411, 1575

\bibitem[{{Bhat} {et~al.}(2004){Bhat}, {Cordes}, {Camilo}, {Nice}, \&
  {Lorimer}}]{bcc04}
{Bhat}, N.~D.~R., {Cordes}, J.~M., {Camilo}, F., {Nice}, D.~J., \& {Lorimer},
  D.~R. 2004, \apj, 605, 759

\bibitem[{{Bhattacharya} \& {van den Heuvel}(1991)}]{bv91}
{Bhattacharya}, D., \& {van den Heuvel}, E.~P.~J. 1991, \physrep, 203, 1

\bibitem[{{Blum} {et~al.}(2003){Blum}, {Ram{\'{\i}}rez}, {Sellgren}, \&
  {Olsen}}]{brs03}
{Blum}, R.~D., {Ram{\'{\i}}rez}, S.~V., {Sellgren}, K., \& {Olsen}, K. 2003,
  \apj, 597, 323

\bibitem[{{Bower} {et~al.}(2006){Bower}, {Goss}, {Falcke}, {Backer}, \&
  {Lithwick}}]{bgf06}
{Bower}, G.~C., {Goss}, W.~M., {Falcke}, H., {Backer}, D.~C., \& {Lithwick}, Y.
  2006, \apjl, 648, L127

\bibitem[{{Boyarsky} {et~al.}(2011){Boyarsky}, {Malyshev}, \&
  {Ruchayskiy}}]{bmr11}
{Boyarsky}, A., {Malyshev}, D., \& {Ruchayskiy}, O. 2011, Physics Letters B,
  705, 165

\bibitem[{{Bridle}(1989)}]{bridle89}
{Bridle}, A.~H. 1989, in Astronomical Society of the Pacific Conference Series,
  Vol.~6, Synthesis Imaging in Radio Astronomy, ed. {R.~A.~Perley,
  F.~R.~Schwab, \& A.~H.~Bridle}, 443--+

\bibitem[{{Cheng} {et~al.}(2006){Cheng}, {Taam}, \& {Wang}}]{ctw06}
{Cheng}, K.~S., {Taam}, R.~E., \& {Wang}, W. 2006, \apj, 641, 427

\bibitem[{{Clark}(1975)}]{clark75}
{Clark}, G.~W. 1975, \apjl, 199, L143

\bibitem[{{Cordes} \& {Lazio}(1997)}]{cl97}
{Cordes}, J.~M., \& {Lazio}, T.~J.~W. 1997, \apj, 475, 557

\bibitem[{{Cordes} \& {Lazio}(2002)}]{cl02}
---. 2002, ArXiv Astrophysics e-prints

\bibitem[{{Crocker} {et~al.}(2011){Crocker}, {Jones}, {Aharonian},
  {et~al.}}]{crocker11}
{Crocker}, R.~M., {Jones}, D.~I., {Aharonian}, F., {et~al.} 2011, \mnras, 413,
  763

\bibitem[{{Deneva}(2010)}]{deneva10}
{Deneva}, I.~S. 2010, PhD thesis, Cornell University

\bibitem[{{Deneva} {et~al.}(2009){Deneva}, {Cordes}, \& {Lazio}}]{dcl09}
{Deneva}, J.~S., {Cordes}, J.~M., \& {Lazio}, T.~J.~W. 2009, \apjl, 702, L177

\bibitem[{{Diehl} {et~al.}(2006){Diehl}, {Halloin}, {Kretschmer},
  {et~al.}}]{diehl06}
{Diehl}, R., {Halloin}, H., {Kretschmer}, K., {et~al.} 2006, \nat, 439, 45

\bibitem[{{Emmering} \& {Chevalier}(1989)}]{ec89}
{Emmering}, R.~T., \& {Chevalier}, R.~A. 1989, \apj, 345, 931

\bibitem[{{Falcke} {et~al.}(1998){Falcke}, {Goss}, {Matsuo},
  {et~al.}}]{falcke98}
{Falcke}, H., {Goss}, W.~M., {Matsuo}, H., {et~al.} 1998, \apj, 499, 731

\bibitem[{{Faucher-Gigu{\`e}re} \& {Kaspi}(2006)}]{fgk06}
{Faucher-Gigu{\`e}re}, C.-A., \& {Kaspi}, V.~M. 2006, \apj, 643, 332

\bibitem[{{Faucher-Gigu{\`e}re} \& {Loeb}(2011)}]{fl11}
{Faucher-Gigu{\`e}re}, C.-A., \& {Loeb}, A. 2011, \mnras, 415, 3951

\bibitem[{{Figer}(2008)}]{figer08}
{Figer}, D.~F. 2008, in IAU Symposium, Vol. 250, IAU Symposium, ed.
  F.~{Bresolin}, P.~A. {Crowther}, \& J.~{Puls}, 247--256

\bibitem[{{Frail} {et~al.}(1994){Frail}, {Diamond}, {Cordes}, \& {van
  Langevelde}}]{fcc94}
{Frail}, D.~A., {Diamond}, P.~J., {Cordes}, J.~M., \& {van Langevelde}, H.~J.
  1994, \apjl, 427, L43

\bibitem[{{Fruchter} \& {Goss}(1990)}]{fg90}
{Fruchter}, A.~S., \& {Goss}, W.~M. 1990, \apjl, 365, L63

\bibitem[{{Gaensler} \& {Slane}(2006)}]{gs06}
{Gaensler}, B.~M., \& {Slane}, P.~O. 2006, \araa, 44, 17

\bibitem[{{Genzel} {et~al.}(2010){Genzel}, {Eisenhauer}, \&
  {Gillessen}}]{geg10}
{Genzel}, R., {Eisenhauer}, F., \& {Gillessen}, S. 2010, Rev. Mod. Phys., 82,
  3121

\bibitem[{{Genzel} {et~al.}(1996){Genzel}, {Thatte}, {Krabbe}, {Kroker}, \&
  {Tacconi-Garman}}]{gtk96}
{Genzel}, R., {Thatte}, N., {Krabbe}, A., {Kroker}, H., \& {Tacconi-Garman},
  L.~E. 1996, \apj, 472, 153

\bibitem[{{Ghez} {et~al.}(2003){Ghez}, {Duch{\^e}ne}, {Matthews},
  {et~al.}}]{gdm03}
{Ghez}, A.~M., {Duch{\^e}ne}, G., {Matthews}, K., {et~al.} 2003, \apjl, 586,
  L127

\bibitem[{{Ghez} {et~al.}(2008){Ghez}, {Salim}, {Weinberg}, \& et~al.}]{gsw08}
{Ghez}, A.~M., {Salim}, S., {Weinberg}, N.~N., \& et~al. 2008, \apj, 689, 1044

\bibitem[{{Gillessen} {et~al.}(2009){Gillessen}, {Eisenhauer}, {Trippe},
  {et~al.}}]{get09}
{Gillessen}, S., {Eisenhauer}, F., {Trippe}, S., {et~al.} 2009, \apj, 692, 1075

\bibitem[{{Heger} {et~al.}(2003){Heger}, {Fryer}, {Woosley}, {Langer}, \&
  {Hartmann}}]{hfw03}
{Heger}, A., {Fryer}, C.~L., {Woosley}, S.~E., {Langer}, N., \& {Hartmann},
  D.~H. 2003, \apj, 591, 288

\bibitem[{{Hooper} \& {Goodenough}(2011)}]{hg11}
{Hooper}, D., \& {Goodenough}, L. 2011, Physics Letters B, 697, 412

\bibitem[{{Johnston} {et~al.}(2006){Johnston}, {Kramer}, {Lorimer},
  {et~al.}}]{jkl06}
{Johnston}, S., {Kramer}, M., {Lorimer}, D.~R., {et~al.} 2006, \mnras, 373, L6

\bibitem[{{Katz}(1975)}]{katz75}
{Katz}, J.~I. 1975, \nat, 253, 698

\bibitem[{{Kramer} {et~al.}(1998){Kramer}, {Xilouris}, {Lorimer},
  {et~al.}}]{kxl98}
{Kramer}, M., {Xilouris}, K.~M., {Lorimer}, D.~R., {et~al.} 1998, \apj, 501,
  270

\bibitem[{{Laguna} \& {Wolszczan}(1997)}]{lw97}
{Laguna}, P., \& {Wolszczan}, A. 1997, \apjl, 486, L27+

\bibitem[{{Lambert} \& {Rickett}(1999)}]{lr99}
{Lambert}, H.~C., \& {Rickett}, B.~J. 1999, \apj, 517, 299

\bibitem[{{Law} {et~al.}(2008){Law}, {Yusef-Zadeh}, {Cotton}, {et~al.}}]{lyc08}
{Law}, C.~J., {Yusef-Zadeh}, F., {Cotton}, W.~D., {et~al.} 2008, \apjs, 177,
  255

\bibitem[{{Lazio} \& {Cordes}(1998)}]{lc98}
{Lazio}, T.~J.~W., \& {Cordes}, J.~M. 1998, \apjs, 118, 201

\bibitem[{{Lazio} \& {Cordes}(2008)}]{lc08}
---. 2008, \apjs, 174, 481

\bibitem[{{Lindqvist} {et~al.}(1992){Lindqvist}, {Habing}, \&
  {Winnberg}}]{lhw92}
{Lindqvist}, M., {Habing}, H.~J., \& {Winnberg}, A. 1992, \aap, 259, 118

\bibitem[{{Liu} {et~al.}(2012){Liu}, {Wex}, {Kramer}, {Cordes}, \&
  {Lazio}}]{lwk12}
{Liu}, K., {Wex}, N., {Kramer}, M., {Cordes}, J.~M., \& {Lazio}, T.~J.~W. 2012,
  \apj, 747, 1

\bibitem[{{L{\"o}hmer} {et~al.}(2001){L{\"o}hmer}, {Kramer}, {Mitra},
  {Lorimer}, \& {Lyne}}]{lkm01}
{L{\"o}hmer}, O., {Kramer}, M., {Mitra}, D., {Lorimer}, D.~R., \& {Lyne}, A.~G.
  2001, \apjl, 562, L157

\bibitem[{{Lorimer} {et~al.}(2006){Lorimer}, {Faulkner}, {Lyne},
  {et~al.}}]{parkesVI2006}
{Lorimer}, D.~R., {Faulkner}, A.~J., {Lyne}, A.~G., {et~al.} 2006, \mnras, 372,
  777

\bibitem[{{Lorimer} {et~al.}(1995){Lorimer}, {Yates}, {Lyne}, \&
  {Gould}}]{lyl95}
{Lorimer}, D.~R., {Yates}, J.~A., {Lyne}, A.~G., \& {Gould}, D.~M. 1995,
  \mnras, 273, 411

\bibitem[{{Lyne} {et~al.}(1998){Lyne}, {Manchester}, {Lorimer},
  {et~al.}}]{parkesII1997}
{Lyne}, A.~G., {Manchester}, R.~N., {Lorimer}, D.~R., {et~al.} 1998, \mnras,
  295, 743

\bibitem[{{Macquart} {et~al.}(2010){Macquart}, {Kanekar}, {Frail}, \&
  {Ransom}}]{mkfr10}
{Macquart}, J.-P., {Kanekar}, N., {Frail}, D.~A., \& {Ransom}, S.~M. 2010,
  \apj, 715, 939

\bibitem[{{Manchester} {et~al.}(2005){Manchester}, {Hobbs}, {Teoh}, \&
  {Hobbs}}]{PSRCAT}
{Manchester}, R.~N., {Hobbs}, G.~B., {Teoh}, A., \& {Hobbs}, M. 2005, \aj, 129,
  1993

\bibitem[{{Maron} {et~al.}(2000){Maron}, {Kijak}, {Kramer}, \&
  {Wielebinski}}]{mkk00}
{Maron}, O., {Kijak}, J., {Kramer}, M., \& {Wielebinski}, R. 2000, \aaps, 147,
  195

\bibitem[{{Miralda-Escud{\'e}} \& {Gould}(2000)}]{meg00}
{Miralda-Escud{\'e}}, J., \& {Gould}, A. 2000, \apj, 545, 847

\bibitem[{{Morris}(1993)}]{morris93}
{Morris}, M. 1993, \apj, 408, 496

\bibitem[{{Muno} {et~al.}(2004){Muno}, {Baganoff}, {Bautz}, {et~al.}}]{mbb04}
{Muno}, M.~P., {Baganoff}, F.~K., {Bautz}, M.~W., {et~al.} 2004, \apj, 613, 326

\bibitem[{{Muno} {et~al.}(2008){Muno}, {Baganoff}, {Brandt}, {Morris}, \&
  {Starck}}]{mbb08}
{Muno}, M.~P., {Baganoff}, F.~K., {Brandt}, W.~N., {Morris}, M.~R., \&
  {Starck}, J.-L. 2008, \apj, 673, 251

\bibitem[{{Muno} {et~al.}(2006){Muno}, {Clark}, {Crowther}, {et~al.}}]{mcc06}
{Muno}, M.~P., {Clark}, J.~S., {Crowther}, P.~A., {et~al.} 2006, \apjl, 636,
  L41

\bibitem[{{Paumard} {et~al.}(2006){Paumard}, {Genzel}, {Martins},
  {et~al.}}]{pgm06}
{Paumard}, T., {Genzel}, R., {Martins}, F., {et~al.} 2006, \apj, 643, 1011

\bibitem[{{Pedlar} {et~al.}(1989){Pedlar}, {Anantharamaiah}, {Ekers},
  {et~al.}}]{pae89}
{Pedlar}, A., {Anantharamaiah}, K.~R., {Ekers}, R.~D., {et~al.} 1989, \apj,
  342, 769

\bibitem[{{Pfahl} \& {Loeb}(2004)}]{pl04}
{Pfahl}, E., \& {Loeb}, A. 2004, \apj, 615, 253

\bibitem[{{Pfahl} {et~al.}(2002){Pfahl}, {Rappaport}, \&
  {Podsiadlowski}}]{prp02}
{Pfahl}, E., {Rappaport}, S., \& {Podsiadlowski}, P. 2002, \apj, 573, 283

\bibitem[{{Pfuhl} {et~al.}(2011){Pfuhl}, {Fritz}, {Zilka}, {et~al.}}]{pfz11}
{Pfuhl}, O., {Fritz}, T.~K., {Zilka}, M., {et~al.} 2011, \apj, 741, 108

\bibitem[{{Reid} \& {Brunthaler}(2004)}]{rb04}
{Reid}, M.~J., \& {Brunthaler}, A. 2004, \apj, 616, 872

\bibitem[{{Roberts}(2004)}]{roberts04}
{Roberts}, M. 2004, The Pulsar Wind Nebula Catalog (March 2005 version),
  \url{http://www.physics.mcgill.ca/~pulsar/pwncat.html}

\bibitem[{{Roy} \& {Rao}(2004)}]{royrao04}
{Roy}, S., \& {Rao}, A.~P. 2004, \mnras, 349, L25

\bibitem[{{Sch{\"o}del} {et~al.}(2009){Sch{\"o}del}, {Merritt}, \&
  {Eckart}}]{sme09}
{Sch{\"o}del}, R., {Merritt}, D., \& {Eckart}, A. 2009, \aap, 502, 91

\bibitem[{{Sch{\"o}del} {et~al.}(2002){Sch{\"o}del}, {Ott}, {Genzel},
  {et~al.}}]{sog02}
{Sch{\"o}del}, R., {Ott}, T., {Genzel}, R., {et~al.} 2002, \nat, 419, 694

\bibitem[{{Skrutskie} {et~al.}(2006){Skrutskie}, {Cutri}, {Stiening},
  {et~al.}}]{2MASS}
{Skrutskie}, M.~F., {Cutri}, R.~M., {Stiening}, R., {et~al.} 2006, \aj, 131,
  1163

\bibitem[{{Tauris} \& {Manchester}(1998)}]{tm98}
{Tauris}, T.~M., \& {Manchester}, R.~N. 1998, \mnras, 298, 625

\bibitem[{{Wang} {et~al.}(2005){Wang}, {Jiang}, \& {Cheng}}]{wjc05}
{Wang}, W., {Jiang}, Z.~J., \& {Cheng}, K.~S. 2005, \mnras, 358, 263

\bibitem[{{Watters} {et~al.}(2009){Watters}, {Romani}, {Weltevrede}, \&
  {Johnston}}]{wrw09}
{Watters}, K.~P., {Romani}, R.~W., {Weltevrede}, P., \& {Johnston}, S. 2009,
  \apj, 695, 1289

\bibitem[{{Wex} \& {Kopeikin}(1999)}]{wk99}
{Wex}, N., \& {Kopeikin}, S.~M. 1999, \apj, 514, 388

\bibitem[{Zhang \& Cheng(1997)}]{zc97}
Zhang, L., \& Cheng, K.~S. 1997, The Astrophysical Journal, 487, 370

\end{thebibliography}

\end{document}